\documentclass[
 aps,physrev,
 reprint,
 groupedaddress,
 amsmath, amssymb,
]{revtex4-2}

\usepackage{braket}
\usepackage{graphicx}
\usepackage{tikz}
\usepackage[version=4]{mhchem}

\usepackage{hyperref}

\DeclareMathOperator{\Tr}{Tr}
\newcommand{\dcolon}{\mathbin{\! : \!}}

\begin{document}

\title{Theory of orientation averaging in X-ray spectroscopies: understanding polarization dependence in a Cartesian tensor approach}

\author{Sihan Zhang}
\affiliation{MaMaSELF$^2$ Course, Université de Montpellier, Place Eugène Bataillon, 34090 Montpellier, France}
\affiliation{ESRF - The European Synchrtron, 71 Avenue des Martyrs, 38043 Grenoble, France}

\author{Oana Bunău}
\affiliation{ESRF - The European Synchrtron, 71 Avenue des Martyrs, 38043 Grenoble, France}

\author{Marius Retegan}
\affiliation{ESRF - The European Synchrtron, 71 Avenue des Martyrs, 38043 Grenoble, France}

\author{Pieter Glatzel}
\affiliation{ESRF - The European Synchrtron, 71 Avenue des Martyrs, 38043 Grenoble, France}

\date{\today}

\begin{abstract}
X-ray absorption spectroscopy (XAS) and resonant inelastic X-ray scattering (RIXS) are powerful probes of electronic structure owing to their chemical and orbital selectivity.
For powder samples, however, interpreting RIXS spectral intensities remains challenging as the measured signal is an average over all orientations.
Existing theoretical treatments rely largely on spherical-tensor formalisms, which often involve complex derivations and case-specific analyses.
Meanwhile, recent advances in quantum-chemistry methods have made the evaluation of transition tensors in Cartesian coordinates both accurate and straightforward.
Here, we present a general theoretical framework that translates Cartesian transition tensors into physically meaningful, orientation-averaged intensities for powder samples.
The formalism allows predicting angular and polarization dependences \textit{ab initio} for both XAS and RIXS and is extendable to other spectroscopies.
The resulting predictions show excellent agreement with RIXS experimental data at the Ce L$_3$ edge.
\end{abstract}

\maketitle

\section{Introduction}

X-ray absorption spectroscopy (XAS) and resonant inelastic X-ray scattering (RIXS) are mature synchrotron radiation techniques 
that offer a unique insight into the electronic structure at the site of the absorbing species \cite{Samak_RIXS_2021, Groot_Resonant_2024}. 
In particular, high-energy-resolution fluorescence-detected XAS (HERFD-XAS), which is in fact a specific cut in the RIXS plane, 
is now an established technique for studying the electronic structure of 3d and 4d transition metals \cite{Glatzel_High_2005,Guo_HERFD_2020,Huang_Resonant_2022,Orduz_L3_2024,Svyazhin_Chemical_2022}, including under \textit{in situ} and \textit{operando} conditions \cite{Beheshti_In_2020, Traulsen_The_2017, Pedersen_Operando_2024, Longo_Dynamic_2022}, as well as in lanthanide and actinide compounds \cite{Zasimov_HERFD_2022, Butorin_High_2016, Amidani_Magnetic_2025, Burrow_Determination_2024, Sundermann_Resonant_2025}.

Beyond the chemical and orbital selectivity inherent to X-ray spectroscopies, 
photon-in - photon-out techniques such as RIXS provide enhanced control through the choice of incident and emitted polarization as well as the scattering geometry. 
However, interpretation of such measurements is often qualitative, commonly relying on comparisons to spectra taken on known references. 
In experimental studies of catalysis and electrochemistry, samples are typically powders \cite{Vedrine_Heterogeneous_2017} providing high surface area and reproducible preparation for spectroscopy and performance measurements. This introduces two major challenges for computing first-principles RIXS spectra of powder samples and for achieving quantitative interpretation of experimental data.

First, it is difficult to reconcile an accurate description of covalency with the inclusion of multiplet effects, arising from electron–electron and electron–core-hole interactions \cite{deGroot_Multiplet_2005, Kotani_Resonant_2001}.
Established electronic-structure calculation methods such as density functional theory (DFT) can effectively describe band structure 
but fail to describe strongly correlated systems, as well as the interaction of the atomic-like states (e.g. $f$) with the localized hole involved in the spectroscopy. 
Conversely, multiplet ligand field theory \cite{Kotani_Resonant_2001} or the single impurity Anderson model \cite{Kotani_Resonant_2001} can accurately capture these interactions, but they require adjustable parameters. The problem can be tackled \textit{ab-initio} by multiconfigurational wavefunction–based quantum-chemistry methods, for example OpenMolcas \cite{Aquilante_Modern_2020} and ORCA \cite{Neese_The_2012}, provided sufficiently large cluster models are employed \cite{Aquilante_Modern_2020, Li_The_2023, Polly_Relativistic_2021, Neese_The_2012}. The codes typically output transition tensors in Cartesian form, which constitutes a further practical consideration motivating our work. Although the Cartesian tensor formulation of X-ray spectroscopy is well established \cite{Joly_Resonant_2012, Bunau_Self_2009}, powder averages for two photon processes are not implemented in calculations packages. Altogether, establishing a general and universal framework that connects Cartesian transition tensors to orientation-averaged intensities has become increasingly important.

Second, because most studied samples are powders, orientation averaging for a general symmetry is required, whereas previous studies of polarization and angular dependence have focused primarily on single crystals 
\cite{Kotani_Theory_2003, Doring_Shake_2004, Gordon_Orientation_2009, Couture_Polarization_2010, Kang_Resolving_2019, Tegomo_Resonant_2022, Krieger_Charge_2022} 
or on highly symmetric systems \cite{Veenendaal_Polarization_2006}. 
Several approaches have been proposed to compute orientation-averaged spectra \cite{Nakazawa_Theory_2000,Juhin_Angular_2014}. 
These methods begin by projecting transition tensors onto a spherical basis and then applying group representation theory. 
Recent developments using this strategy provide a full solution, i.e. orientation averages are expressed as linear combinations of fundamental spectra obtained within the electric dipole (E1E1) approximation for all point-symmetry groups \cite{ Burrow_Angular_2026, Tagliavini_Polarization_2025}. However, these approaches cannot easily be extended to higher-order multipole contributions, due to the large number of terms involved ($3^6$ for E2E1 compared to $3^4$ for E1E1). 

This work presents a general framework for calculating orientation-averaged transition intensities of linearly polarized X-rays and provides a unified description of angular and polarization dependence in XAS and RIXS processes.

Section II introduces the theoretical framework and the key mathematical foundations.  We explicitly separate the experimental geometry (polarizations and wavevector orientations) from the sample geometry.

Section III derives the orientation-averaged expressions for XAS and RIXS in terms of Cartesian tensors within a unified formalism. For scattering processes, we consider the particular case of  the fundamental in-plane ($\pi$) and out-of-plane ($\sigma$) polarization configurations and show an analogy with the classical diffraction limit. Some of the results are original - most notably the powder averages for scattering processes - while others, such as the powder average of quadrupolar absorption or the lack of dipole-quadrupole interference contribution in powders, are already known and serve to validate the robustness of our theoretical framework.

Section IV presents an application to experimentally measured core-to-core 2$p$3$d$ and valence-to-core RIXS at the cerium L$_3$ edge in \ce{CeO2}.

A detailed analytical description of the results and the equivalence of the Cartesian and spherical tensor approaches are shown in the appendix.

\section{Model description and mathematical background}
\subsection{XAS and RIXS cross sections}
\label{sec:crosssec}
\begin{table}[t]
    \caption{Notation used throughout the paper.}
    \begin{ruledtabular}
    \begin{tabular}{ll}
        Symbol & Meaning \\ 
        \hline
        $\Re(a)$ & Real part \\
        $\Im(a)$ & Imaginary part \\
        $\Tr(\mathbf{A})$ & Trace \\
        $\mathbf{A}^*,\, a^*$ & Complex conjugate \\
        $\mathbf{A}^\mathrm{T}$ & Transpose \\
        $\mathbf{A}^\dagger$ & Hermitian adjoint \\
        $\mathbf{A}\dcolon\mathbf{B}$ & Tensor contraction: $\sum_{ijk\cdots} A_{ijk\cdots} B_{ijk\cdots}$ \\
        $\mathbf{A}\otimes\mathbf{B}$ & Tensor product \\
        $\mathbf{D},\, D_i$ & Rank-1 dipole (E1) transition tensor\\
        $\boldsymbol{\mathcal{D}},\, \mathcal{D}_{ij}$ & $\boldsymbol{\mathcal{D}}=\mathbf{D}^*\otimes \mathbf{D}$, first order rank-2 (E1) tensor \\
        $\mathbf{Q},\, Q_{ij}$ & Rank-2 quadrupole (E2) transition tensor\\
        $\boldsymbol{\mathcal{R}},\, \mathcal{R}_{ijk}$ & $\boldsymbol{\mathcal{R}}=\mathbf{D}^*\otimes \mathbf{Q}$ rank-3 E1E2 interference tensor\\
        $\boldsymbol{\mathcal{Q}},\, \mathcal{Q}_{ijkl}$ & $\boldsymbol{\mathcal{Q}}=\mathbf{Q}^*\otimes \mathbf{Q}$, first order rank-4 (E2) tensor \\
        $\mathbf{M},\, M_{ij}$ & second order rank-2 (E1E1) tensor \\
        $\mathbf{S},\, S_{ijk}$ & second order rank-3 (E2E1) tensor
    \end{tabular}
    \end{ruledtabular}
    \label{tab:notations}
\end{table}

The transition operator form used in this work contains terms of electrical origin only \cite{Sakurai_Advanced_1967, Bernadotte_Origin_2012}:
\begin{equation}
    \hat{\mathbf{T}} =  \boldsymbol{\varepsilon}\cdot\boldsymbol{r} + 
    \frac{i}{2}(\boldsymbol{\varepsilon} \cdot \boldsymbol{r}) (\boldsymbol{k} \cdot \boldsymbol{r}) +  \mathcal{O}( \boldsymbol{k} \cdot \boldsymbol{r})^2\label{eq:TansitionOperator}
\end{equation}
where the first term corresponds to the electric dipole (E1) transition, the second term corresponds to the electric quadrupole (E2) transition and the higher expansion terms are neglected. $\boldsymbol r$ is the position operator and $k$ the wavevector of the photon.

XAS is a one-photon absorption process and is therefore described to first order in perturbation theory. Using Fermi’s golden rule the XAS cross section reads
\begin{equation}
    \Sigma_{\mathrm{XAS}}(\hbar\omega) = 4\pi^2\alpha_{\mathrm{fs}}\hbar\omega \sum_{\ket{f}} |\bra{f}\hat{\mathbf{T}}\ket{g}|^2 \delta(E_f-E_g-\hbar\omega) \label{eq:Fermi}
\end{equation}
in units of $[L^2]$ with $\alpha_{\mathrm{fs}}$ the fine-structure constant    \cite{Sakurai_Advanced_1967}. It describes the transition from an initial electronic state $\ket{g}$ of energy $E_g$ to final state $\ket{f}$ of energy $E_f$ upon absorbing a photon of energy $\hbar\omega$.

RIXS is a two-step scattering process (absorption followed by emission) described in the second order of perturbation theory \cite{Sakurai_Advanced_1967, Ament_Resonant_2011, Hassing_The_2004, Juhin_Angular_2014} with cross section:
\begin{eqnarray}
     \Sigma_{\mathrm{RIXS}}(\omega_i, \omega_o) = 
\frac{\alpha_{\mathrm{fs}}^2}{\hbar^2c^2}\frac{\omega_{o}}{\omega_{i}} \sum_{\ket{f}} \delta(E_f-E_g-\hbar(\omega_i-\omega_o)) \nonumber \\
\times \left|\sum_{\ket{n}} (E_f-E_n) \cdot (E_n - E_g) \frac{\bra{f}\hat{\mathbf{T}}_{o}^\dagger\ket{n} \bra{n}\hat{\mathbf{T}}_{i}\ket{g}}{\hbar\omega_{i} - (E_n-E_g) +i \Gamma_n}\right|^2 
    \label{eq:KramersHeisenberg} 
\end{eqnarray}
in units of $[L^2]/[E]$ with $c$ the speed of light. 
Here $\hat{\mathbf{T}}_i$ ($\hat{\mathbf{T}}_o$) is the transition operator corresponding to the absorption (emission) process of the incoming (outgoing) photon of energy $\hbar\omega_i$ ($\hbar\omega_o$) and $\ket{n}$ denotes the intermediate states of energy $E_n$. 

In this manuscript we explicitly take the convolution function as the commonly used Lorentzian form, and distinguish between the total scattering section $\Sigma$ and $\sigma$, the latter corresponding to a single final state. The following relationship holds for both XAS and RIXS:
\begin{equation}
\begin{aligned}
\Sigma &= C \sum_{\ket{f}} \frac{1}{\pi}
\frac{\Gamma_f}{(E_f - E_g - \hbar \Omega)^2 + \Gamma_f^2}
\, \sigma(E_f, E_g) \\
\Omega &=
\begin{cases}
\omega & \text{(XAS)} \\
\omega_i - \omega_o & \text{(RIXS)}
\end{cases}
\end{aligned}
\label{eq:E_conv}
\end{equation}
with $C$ the numerical prefactor in Eqs. \ref{eq:Fermi} and \ref{eq:KramersHeisenberg}. The Dirac delta was replaced by a Lorentzian function of full width at half maximum 2$\Gamma_f$. Note that prefactors in Eqs. \ref{eq:Fermi} and \ref{eq:KramersHeisenberg} are specific to the chosen transition operator form in Eq. \ref{eq:TansitionOperator}.

\subsection{Orientation averaging and tensor-integral reduction}

\label{sec:average}
In their standard form, Eqs.~\ref{eq:Fermi} and \ref{eq:KramersHeisenberg} are not well suited for analyzing orientation dependence.
To address this, we seek an alternative representation that cleanly separates the angular contributions originating from the experimental geometry
(i.e., the orientations of the polarization vectors and wavevectors) from those inherent to the sample, encoded in the transition matrix elements.
The latter obey the symmetry constraints of the point group of the absorbing atom and therefore possess their own intrinsic angular dependence.
To make this separation explicit, the cross sections can be recast as a series of tensor contractions:
\begin{eqnarray}
    \sigma &=& |(\boldsymbol{v}_1 \cdot \mathbf{D}_1) \cdots (\boldsymbol{v}_k \cdot \mathbf{D}_k)|^2 \nonumber\\ 
    &=& (\boldsymbol{v}_{1}^* \otimes \cdots \otimes \boldsymbol{v}_{k}^* \otimes \boldsymbol{v}_1 \otimes \cdots \otimes \boldsymbol{v}_k) \nonumber\\
    && \dcolon\,(\mathbf{D}_{1}^* \otimes \cdots \otimes \mathbf{D}_{k}^* \otimes \mathbf{D}_1 \otimes \cdots \otimes \mathbf{D}_k)
    \label{eq:sigmaTensor}
\end{eqnarray}
where we used the notations explicitly stated in Table~\ref{tab:notations} and $\boldsymbol{v}$ denotes vectors such as polarization $\boldsymbol{\varepsilon}$ and wavevector $\boldsymbol{k}$.

Thanks to Eq.~\ref{eq:sigmaTensor} the cross section at any symmetry-equivalent site can be obtained simply by applying the corresponding transformation $\mathbf{R}$ to the transition tensors.
\begin{equation}
     \sigma^{(\mathbf{R})} = (\boldsymbol{v}_{1}^* \otimes \cdots \otimes \boldsymbol{v}_k) \dcolon((\mathbf{R}\mathbf{D}_{1})^* \otimes \cdots \otimes \mathbf{R}\mathbf{D}_k) \label{eq:sigmaR}   
\end{equation}

For non-magnetic and non-chiral materials, the cross section of a powder sample is obtained by averaging over all possible crystallite orientations, i.e., over the SO(3) rotation group:
\begin{equation}
    \braket{\sigma}_\mathbf{R} = \int_{SO(3)} \sigma^{(\mathbf{R})} \,d\mathbf{R} \label{eq:powderAverage}
\end{equation}
with $\mathbf{R}$ an orthogonal rotation matrix and $d\mathbf{R}$ the normalized Haar measure on SO(3). 

Whereas for a single crystal one must explicitly sum the contributions from all symmetry-equivalent sites of the absorbing atom, 
in a powder these contributions are automatically included by the rotational averaging.
Indeed, the isotropic components associated with two symmetry-related sites connected by  rotation $g$ are identical. As a consequence of the invariance property of the Haar integral:
\begin{equation}
\int_{SO(3)} \sigma^{(\mathbf{R})} \,d\mathbf{R} \equiv \int_{SO(3)} \sigma^{(\mathbf{Rg})} \,d\mathbf{R} \label{eq:SO3int}
\end{equation}
where $\sigma^{(\mathbf{Rg})}$ is the result of applying $g$ to the transition tensors.
Thus, once the transition tensor is constructed for the minimal set of sites that generate all others via real-space symmetry operations, 
no additional accounting for symmetry-equivalent atomic positions is required in the powder average. For samples containing non-equivalent atomic sites (in the sense of this work, including those related by improper symmetry operations), the orientational averages must be evaluated separately for each site and subsequently summed.

Direct evaluation of the integral in Eq.~\ref{eq:powderAverage} is generally impractical because the number of terms grows rapidly with the tensor rank.
A more tractable strategy is to exploit combinatorial methods.
In particular, it is known that the rotational average of any even-rank Cartesian tensor over SO(3)
can be written as a linear combination of products of Kronecker deltas, 
with coefficients determined solely by combinatorial factors \cite{Rashid_Linear_2011, Ee_Combinatorics_2017}.

To illustrate the procedure, first introduced in Ref. \cite{Andrews_On_1977}, consider a rank-4 Cartesian tensor $\boldsymbol{\mathcal{Q}}$.
Its rotational average admits an expansion in the basis formed by all independent products of Kronecker deltas:
\begin{equation}
    \braket{\mathcal{Q}}_{ijkl} = \alpha\,A_{ijkl} + \beta\,B_{ijkl} + \gamma\,C_{ijkl} \label{eq:rank4Average}
\end{equation}
where:
\begin{eqnarray}
    A_{ijkl} &=& \delta_{ik}\delta_{jl} \nonumber\\
    B_{ijkl} &=& \delta_{ij}\delta_{kl} \nonumber\\
    C_{ijkl} &=& \delta_{il}\delta_{jk}
\end{eqnarray}
The expansion coefficients $\alpha$, $\beta$, and $\gamma$ can be obtained by solving the contraction equation, which means the averaged $\braket{\boldsymbol{\mathcal{Q}}}$ and the original tensor $\boldsymbol{\mathcal{Q}}$ should have the same projection $\braket{\mathcal{Q}}_{ijkl}$ on the deltas basis $A_{ijkl}$, $B_{ijkl}$ and $C_{ijkl}$:
\begin{equation}
    \begin{bmatrix}
        \mathbf{A}\dcolon\braket{\boldsymbol{\mathcal{Q}}} \\ \mathbf{B}\dcolon\braket{\boldsymbol{\mathcal{Q}}} \\ \mathbf{C}\dcolon\braket{\boldsymbol{\mathcal{Q}}}
    \end{bmatrix} = 
    \begin{bmatrix}
        \mathbf{A}\dcolon\mathbf{A} & \mathbf{A}\dcolon\mathbf{B} & \mathbf{A}\dcolon\mathbf{C} \\
        \mathbf{B}\dcolon\mathbf{A} & \mathbf{B}\dcolon\mathbf{B} & \mathbf{B}\dcolon\mathbf{C} \\
        \mathbf{C}\dcolon\mathbf{A} & \mathbf{C}\dcolon\mathbf{B} & \mathbf{C}\dcolon\mathbf{C}
    \end{bmatrix}
    \begin{bmatrix}
        \alpha \\ \beta \\ \gamma
    \end{bmatrix} = 
    \begin{bmatrix}
        \mathbf{A}\dcolon\boldsymbol{\mathcal{Q}} \\ \mathbf{B}\dcolon\boldsymbol{\mathcal{Q}} \\ \mathbf{C}\dcolon\boldsymbol{\mathcal{Q}}
    \end{bmatrix}
\end{equation}
The middle matrix is the Gram matrix. In three dimensions, the diagonal elements are 9 and the off-diagonal elements are 3.

Subsequently, by using the inverse of the above, we obtained the parameter $\alpha$, $\beta$ and $\gamma$:
\begin{equation}
    \begin{bmatrix}
        \alpha \\ \beta \\ \gamma
    \end{bmatrix} = 
    \frac{1}{30}\begin{bmatrix}
        4 & -1 & -1 \\
        -1 & 4 & -1 \\
        -1 & -1 & 4 \\
    \end{bmatrix}
    \sum_{ijkl}
    \begin{bmatrix}
        \delta_{ik}\delta_{jl} \\
        \delta_{ij}\delta_{kl} \\
        \delta_{il}\delta_{jk}
    \end{bmatrix} \mathcal{Q}_{ijkl} \label{eq:rank4Average_coef}
\end{equation}
By plugging parameters back to Eq. \ref{eq:rank4Average},
we can obtain the averaged tensor projection.

Formally the same procedure can be applied to obtain the rotational average of a rank-2 tensor $\boldsymbol{\mathcal{D}}$:
\begin{equation}
    \braket{\mathcal{D}}_{ij} = \alpha\,\delta_{ij}  \label{eq:rank2Average}
\end{equation}
where the Gram matrix reduces to a scalar $\alpha$ given by:
\begin{equation}
    \alpha = \frac{1}{3}\sum_{ij}\delta_{ij}\mathcal{D}_{ij} = \frac{1}{3} \Tr(\boldsymbol{\mathcal{D}})\label{eq:rank2Average_coef}
\end{equation}

For higher-order tensors (rank 6 and above), the associated Gram matrices have been computed by D. L. Andrews \cite{Andrews_On_1977, Andrews_Eighth_1981}. These results will be used directly in the calculations that follow.

\section{Orientation averages for X-ray spectroscopy}

The rotational averages $\braket{\boldsymbol{\mathcal{D}}}$ and $\braket{\boldsymbol{\mathcal{Q}}}$ involve only the transition tensors.
For the evaluation of spectroscopic observables, however, the experimental geometry must be included as in Eq.~\ref{eq:sigmaTensor}.

\subsection{XAS}
We start with the simplest case, namely the powder average of the E1 XAS term.
Using Eqs.~\ref{eq:rank2Average} and \ref{eq:rank2Average_coef}, we recover the well-known result:
\begin{eqnarray}
     \braket{\sigma_\mathrm{XAS}^\mathrm{E1}} &=& \sum_{ij} \varepsilon_i^* \braket{\mathcal{D}}_{ij} \varepsilon_j  \nonumber \\
     &=& \frac{1}{3}|\boldsymbol{\varepsilon}|^2 \Tr{(\boldsymbol{\mathcal{D}})} 
     = \frac{1}{3}|\boldsymbol{\varepsilon}|^2 |\mathbf{D}|^2 \label{eq:sigmaD}
\end{eqnarray}
where, formally, one may work either with the rank-2 E1 tensor $\boldsymbol{\mathcal{D}} = \mathbf{D}^* \otimes \mathbf{D}$ or directly with the rank-1 E1 transition vector $\mathbf{D}$ defined as $\bra{f}\boldsymbol{r}\ket{g}$. 

The E2 XAS term is associated to a rank-4 tensor. Its orientation average is:
\begin{eqnarray}
     \braket{\sigma_\mathrm{XAS}^\mathrm{E2}} &=& \sum_{ijkl} \varepsilon_i^* k_j^* \braket{\mathcal{Q}}_{ijkl} \varepsilon_k k_l
\end{eqnarray}
based on the definition in Eq.~\ref{eq:rank4Average}. Eventually from Eq.~\ref{eq:rank4Average_coef}:
\begin{eqnarray}
    \braket{\sigma_\mathrm{XAS}^\mathrm{E2}} &=& \frac{1}{30} \begin{bmatrix}
        |\boldsymbol{\varepsilon}|^2|\boldsymbol{k}|^2 \\
        |\boldsymbol{\varepsilon}\cdot\boldsymbol{k}|^2 \\
        |\boldsymbol{\varepsilon}^*\cdot\boldsymbol{k}|^2
    \end{bmatrix}^\mathrm{T}
    \begin{bmatrix}
        4 & -1 & -1 \\
        -1 & 4 & -1 \\
        -1 & -1 & 4 \\
    \end{bmatrix} \sum_{ij}
    \begin{bmatrix}
        \mathcal{Q}_{ijij} \\
        \mathcal{Q}_{iijj} \\
        \mathcal{Q}_{ijji}
    \end{bmatrix} \nonumber
\end{eqnarray}

For convenience we choose to work with the rank-2 E2 tensor $\mathbf{Q}=\bra{f}\boldsymbol{r}\otimes\boldsymbol{r}\ket{g}$, symmetric by construction ($Q_{ij} = Q_{ji}$). With this the E2 XAS isotropic signal reads:
\begin{eqnarray}
   \braket{\sigma_\mathrm{XAS}^\mathrm{E2}} &=& \frac{1}{30} \begin{bmatrix}
        |\boldsymbol{\varepsilon}|^2|\boldsymbol{k}|^2 \\
        |\boldsymbol{\varepsilon}\cdot\boldsymbol{k}|^2 \\
        |\boldsymbol{\varepsilon}^*\cdot\boldsymbol{k}|^2
    \end{bmatrix}^\mathrm{T}
    \begin{bmatrix}
        4 & -1 & -1 \\
        -1 & 4 & -1 \\
        -1 & -1 & 4 \\
    \end{bmatrix} \sum_{ij}  
    \begin{bmatrix}
        Q_{ij}^*Q_{ij} \\
        Q_{ii}^*Q_{jj} \\
        Q_{ij}^*Q_{ji}  
    \end{bmatrix} \nonumber 
\end{eqnarray}
Since in first order processes the polarization $\boldsymbol{\varepsilon}$ is perpendicular to the wavevector $\boldsymbol{k}$, the above reduces to: 
\begin{eqnarray}
\braket{\sigma_\mathrm{XAS}^\mathrm{E2}}    &=& \frac{1}{30} |\boldsymbol{\varepsilon}|^2|\boldsymbol{k}|^2 \times  \\
&\times& \left(4\sum_{ij} Q_{ij}^*Q_{ij} - \sum_{ij} Q_{ii}^*Q_{jj} - \sum_{ij} Q_{ij}^*Q_{ji} \right) \nonumber
\end{eqnarray}
We use $\mathbf{Q}$'s symmetry property to further simplify to:
\begin{eqnarray}
    \braket{\sigma_\mathrm{XAS}^\mathrm{E2}}  &=& \frac{1}{30} |\boldsymbol{\varepsilon}|^2|\boldsymbol{k}|^2  \left(3\Tr(\mathbf{Q}^\dagger\mathbf{Q}) - |\Tr(\mathbf{Q})|^2 \right) \nonumber\label{eq:sigmaQ1}       
\end{eqnarray}
with $\Tr(\mathbf{Q}^\dagger\mathbf{Q})=\sum_{ij}Q_{ij}^*Q_{ij}$,
$|\Tr(\mathbf{Q})|^2 =  \sum_{ij} Q_{ii}^* Q_{jj} $.

With explicit cartesian indices:
\begin{eqnarray}
    \braket{\sigma_\mathrm{XAS}^\mathrm{E2}} 
    &=& \frac{|\boldsymbol{\varepsilon}|^2|\boldsymbol{k}|^2}{15}\left(|Q_{xx}|^2 + 3|Q_{xy}|^2 - \Re(Q_{xx}^*Q_{yy})\right)  \nonumber \\ 
    &+& \text{cyclic permutations}
    \label{eq:sigmaQ2}
\end{eqnarray}
The result above is identical to the one in reference \cite{Brouder_Site_2008}, provided we expand on the real valued rank-4 tensor defined as  $\mathcal{Q}_{ijkl}^\prime = (Q_{ij}^*Q_{kl} + Q_{ij}Q_{kl}^*)/2$:
\begin{eqnarray}
    \braket{\sigma_\mathrm{XAS}^\mathrm{E2}} 
    &=& \frac{|\boldsymbol{\varepsilon}|^2|\boldsymbol{k}|^2}{15}(\mathcal{Q}_{xxxx}^\prime + 3\mathcal{Q}_{xyxy}^\prime - \mathcal{Q}_{xxyy}^\prime)  \nonumber \\ 
    &+& \text{cyclic permutations}
    \label{eq:sigmaQ4}
\end{eqnarray}

Equations~\ref{eq:sigmaD} and \ref{eq:sigmaQ2} make it evident that the E1 and E2 contributions to XAS in a powder sample exhibit no dependence on either wavevector orientation or polarization.

Under specific conditions, E1 and E2 transitions may interfere:
\begin{eqnarray}
    \sigma_{\mathrm{XAS}}^{\mathrm{E1+E2}} &=& |\bra{f}\boldsymbol{\varepsilon}\cdot\boldsymbol{r}+\frac{i}{2}(\boldsymbol{\varepsilon}\cdot\boldsymbol{r})(\boldsymbol{r}\cdot\boldsymbol{k})\ket{g}|^2 \\
    &=& \sigma_{\mathrm{XAS}}^{\mathrm{E1}} + \frac{1}{4}\sigma_{\mathrm{XAS}}^{\mathrm{E2}} \nonumber\\
    &&- \Im(\bra{g}(\boldsymbol{\varepsilon}\cdot\boldsymbol{r})^*\ket{f}\bra{f}(\boldsymbol{\varepsilon}\cdot\boldsymbol{r})(\boldsymbol{r}\cdot\boldsymbol{k})\ket{g}) \nonumber
\end{eqnarray}
The last term corresponds to the E1E2 interference contribution, denoted $\sigma_{\mathrm{XAS}}^{\mathrm{E1E2}}$. It involves the $\boldsymbol{\mathcal{R}}=\mathbf{D}^*\otimes \mathbf{Q}$ rank 3 tensor where  $\mathcal{R}_{ijk}=\bra{g}r_i\ket{f}\bra{f} r_j r_k\ket{g}$. Subsequently the interference term can be written as :
\begin{eqnarray}
    \sigma_{\mathrm{XAS}}^{\mathrm{E1E2}}
    &=& - \Im\left(\sum_{ijk}\varepsilon^*_i\varepsilon_j k_k \mathcal{R}_{ijk}\right)
\end{eqnarray}

The orientation average of odd ranked tensors may be evaluated using a procedure analogous to that introduced in Section~\ref{sec:average}, equally developed in Ref. \cite{Andrews_On_1977}. For rank 3 tensors, the only isotropic basis is provided by the antisymmetric Levi-Civita symbol $\epsilon_{ijk}$. Accordingly, the orientation average can be expanded as $\braket{\mathcal{R}}
_{ijk} = \beta \epsilon_{ijk}$, where $\beta =\frac{1}{6} \sum_{ijk} \epsilon_{ijk} \mathcal{R}_{ijk}$. This yields
\begin{eqnarray}
    \braket{\sigma_{\mathrm{XAS}}^{\mathrm{E1E2}}} 
    = - \Im\frac{1}{6}\left(\sum_{ijk}\epsilon_{ijk}\varepsilon^*_i\varepsilon_j k_k\right)\left(\sum_{ijk}\epsilon_{ijk}\mathcal{R}_{ijk}\right) \nonumber \\ \label{eq:avgE1E2interf}
\end{eqnarray}

Here, $j$ and $k$ are the two indices of the symmetric E2 tensor $\mathbf{Q}$. The second term of the product in Eq.~\ref{eq:avgE1E2interf} is a Levi-Civita sum with two switchable indexes, so it must be zero: 
$$\braket{R}_{ijk} = \braket{R}_{ikj} , \epsilon_{ijk} = -\epsilon_{ikj} \Rightarrow \beta = 0$$
This is consistent to the known result,  namely that the E1E2 interference has no isotropic component and therefore does not contribute to the absorption cross section in powder samples \cite{Peacock_Natural_2001, Carra_Xray_2003}.

\subsection{RIXS}
\subsubsection{General formalism}

The E1E1 RIXS term is very convenient to express in terms of rank-2 tensors, in particular: 
\begin{eqnarray}
    M_{ij} &=& \sum_{n} (E_f-E_n) \cdot (E_n - E_g)\frac{\bra{n}r_i\ket{g}\cdot\bra{f}r_j\ket{n}}{\hbar\omega-(E_n-E_g)+i\Gamma_n} \nonumber \\
    \label{eq:defM}
\end{eqnarray}
where $\hbar\omega$ is the incoming photon energy. The E1E1 RIXS tensor is rank-4 and relates to 
$\mathbf{M}^* \otimes \mathbf{M}$. Its orientation average is:
\begin{eqnarray} \braket{\sigma_\mathrm{RIXS}^\mathrm{E1E1}} &=& \sum_{ijkl} \varepsilon_{in,i}^* \varepsilon_{out,j}\ \braket{M_{ij}^*M_{kl}} \
\varepsilon_{in,k} \varepsilon_{out,l}^*
\end{eqnarray}
with $\boldsymbol{\varepsilon}_{in}$ ($\boldsymbol{\varepsilon}_{out}$) the incoming (outgoing) polarization, denoted simply by  $\boldsymbol{\varepsilon}_{i}$ ($\boldsymbol{\varepsilon}_{o}$) when there is no risk of confusion. As explained in section \ref{sec:crosssec}, $\braket{\sigma_\mathrm{RIXS}^\mathrm{E1E1}}$ corresponds to a single final state and omits prefactors. In practice, the denominator in Eq. \ref{eq:defM} is replaced by a Lorentzian function of width 2$\Gamma_n$.

From this point the derivation is similar to the one of E2 XAS:
\begin{eqnarray}
    \braket{\sigma_\mathrm{RIXS}^\mathrm{E1E1}} &=& \frac{1}{30} \begin{bmatrix}
        
        |\boldsymbol{\varepsilon}_o|^2 |\boldsymbol{\varepsilon}_i|^2 \\
        |\boldsymbol{\varepsilon}_o^*\cdot\boldsymbol{\varepsilon}_i|^2 \\
        |\boldsymbol{\varepsilon}_o\cdot\boldsymbol{\varepsilon}_i|^2 
    \end{bmatrix}^\mathrm{T}
    \begin{bmatrix}
        4 & -1 & -1 \\
        -1 & 4 & -1 \\
        -1 & -1 & 4 \\
        \end{bmatrix} \sum_{ij}  
    \begin{bmatrix}
        M_{ij}^*M_{ij} \\
        M_{ii}^*M_{jj} \\
        M_{ij}^*M_{ji}  
    \end{bmatrix} \nonumber
\end{eqnarray}
It can be easily shown that:
\begin{eqnarray}
\Tr(\mathbf{M}^\dagger\mathbf{M}) &=& \sum_{ij}M_{ij}^\dagger M_{ji}  = \sum_{ij}M_{ij}^*M_{ij}\nonumber \\
\Tr(\mathbf{M}^*\mathbf{M}) &=&  \sum_{ij}M_{ij}^*M_{ji} \nonumber \\
|\Tr(\mathbf{M})|^2 &=&  \sum_{ij} M_{ii}^* M_{jj}
\end{eqnarray}
with which:
\begin{eqnarray}
    \braket{\sigma_\mathrm{RIXS}^\mathrm{E1E1}} &=& \frac{1}{30} \begin{bmatrix}
        |\boldsymbol{\varepsilon}_o|^2 |\boldsymbol{\varepsilon}_i|^2 \\ 
        |\boldsymbol{\varepsilon}_o^*\cdot\boldsymbol{\varepsilon}_i|^2 \\
        |\boldsymbol{\varepsilon}_o\cdot\boldsymbol{\varepsilon}_i|^2
    \end{bmatrix}^\mathrm{T} \times \nonumber \\
    &\times&
    \begin{bmatrix}
        4 & -1 & -1 \\
        -1 & 4 & -1 \\
        -1 & -1 & 4 \\
        \end{bmatrix}  
    \begin{bmatrix}
        \Tr(\mathbf{M^\dagger\mathbf{M}}) \\
        |\Tr(\mathbf{M})|^2  \\
        \Tr(\mathbf{M^*\mathbf{M}})
    \end{bmatrix} 
    \label{eq:sigmaDD} \nonumber\\
\end{eqnarray}

In photon-in photon-out (second order) processes the incident and scattered polarization vectors, $\boldsymbol{\varepsilon}_i$ and $\boldsymbol{\varepsilon}_o$, are not constrained to be orthogonal. This lack of orthogonality is the origin of the angular and polarization dependence that appears in RIXS measurements on powders. This contrasts with the XAS E2 case, where no angular dependence arises because the wavevector is always perpendicular to the polarization. Moreover, Eq.~\ref{eq:sigmaDD} highlights an additional important consequence: when the two polarization vectors are mutually perpendicular, the intensity becomes independent of both angle and polarization. A similar conclusion was reached in Ref.~\cite{Juhin_Angular_2014}.

The intensity for any higher multipole order can be evaluated in a similar way, with the Gram matrix taken from \cite{Andrews_On_1977, Andrews_Eighth_1981}. For higher orders the advantage of 1- and 2-index notations becomes obvious. Let us consider the E2E1 RIXS (E2 in, E1 out), described by the rank-6 tensor:
\begin{eqnarray}
\sigma_{\mathrm{RIXS}}^{\mathrm{E2E1}} = 
\left|\sum_{ijk} \boldsymbol{\varepsilon}_{in,i} k_{in,j} \boldsymbol{\varepsilon}_{out,k}^* S_{ijk} \right|^2 \nonumber
\end{eqnarray}
with the rank-3 tensor $\mathbf{S}$ defined as
\begin{eqnarray}
    S_{ijk} &=& \sum_{n} \frac{\bra{n}r_ir_j\ket{g}\cdot\bra{f}r_k\ket{n}}{\hbar\omega-(E_n-E_g)+i\Gamma_n}
    \label{eq:defS}
\end{eqnarray}
The orientation average of the E2E1 RIXS term is:
\begin{eqnarray}
    &&\braket{\sigma_{\mathrm{RIXS}}^{\mathrm{E2E1}}} = \frac{1}{210} 
    \begin{bmatrix}
        |\boldsymbol{\varepsilon}_o|^2 |\boldsymbol{\varepsilon}_i|^2 |\boldsymbol{k}_i|^2 \\
        |\boldsymbol{\varepsilon}_o \cdot \boldsymbol{k}_i|^2 |\boldsymbol{\varepsilon}_i|^2 \\
        |\boldsymbol{\varepsilon}_o^* \cdot \boldsymbol{\varepsilon}_i|^2 |\boldsymbol{k}_i|^2 \\
        |\boldsymbol{\varepsilon}_o \cdot \boldsymbol{\varepsilon}_i|^2 |\boldsymbol{k}_i|^2 
    \end{bmatrix}^{\mathrm{T}} \times \nonumber\\
    && 
    \begin{bmatrix}
        11 & -6 & -6 & -5 &   8 \\
        -6 &  9 &  9 &  4 & -12 \\
        -3 &  8 &  1 &  2 &  -6 \\
        -3 &  1 &  8 &  2 &  -6 
    \end{bmatrix} \nonumber\\
    && \times\sum_{ii'jj'kk'}
    \begin{bmatrix}

        \delta_{ii'}\delta_{jj'}\delta_{kk'} \\ \delta_{ii'}\delta_{jk}\delta_{j'k'} \\
        \delta_{ii'}\delta_{jk'}\delta_{j'k} \\
        \delta_{ij}\delta_{i'j'}\delta_{kk'} \\ \delta_{ij}\delta_{i'k}\delta_{j'k'} \Re  
    \end{bmatrix}  S^*_{i'j'k'}S_{ijk}
    \label{eq:sigmaDQ}
\end{eqnarray}

Eq.~\ref{eq:sigmaDQ} maps the space of rank-6 isotropic tensors (dimension $6!/(2^3\cdot3!) = 15$, i.e. number of independent ways to pair 6 indices \cite{Andrews_On_1977}) onto a lower-dimensional space. First, redundancy was removed by considering the orthogonality of $\boldsymbol{\varepsilon}_i$ and $k_i$. Second, more simplification can be achieved by exploiting the $S_{ijk} = S_{jik}$ symmetry. Third, we used complex conjugation. The full proof is to be found in the Supplementary Information. 

An interesting observation is that the rank of the reduced $4\times 5$ matrix is 3, implying that only three linearly independent (fundamental) spectra contribute to E2E1 RIXS for an isotropic sample. This result is consistent with Ref.~\cite{Juhin_Angular_2014}.

The orientation averaged E1E2 RIXS can be constructed in a similar way. The corresponding result consists in interchanging $\varepsilon_i$ and $\varepsilon_o$ in Eq.~\ref{eq:sigmaDQ} above.

\subsubsection{Projections onto polarization channels}

In the previous section, we derived the general expression for the RIXS scattering cross section corresponding to arbitrary incident and emitted polarizations. In practice, however, this formulation has limited direct applicability, since most experiments employ linearly polarized incident light and polarization-insensitive detection. Consequently, the intensity must be projected onto the polarization configuration dictated by the experimental geometry.

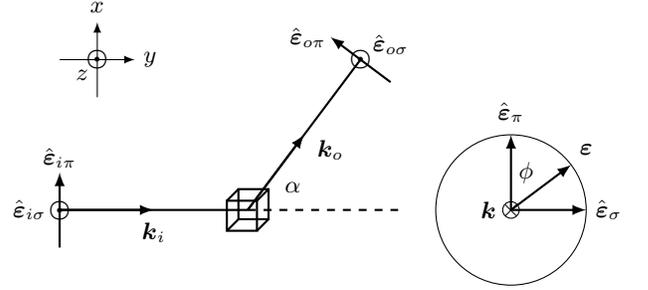
\begin{figure}[!htbp]
    \centering
    \begin{tikzpicture}
        \draw[->, >=latex, thick] (-2.5, 0, 0) -- (-1.25, 0, 0); \draw[-, thick] (-2.5, 0, 0) -- (0, 0, 0); \node at (-1.25, -0.3, 0) {$\boldsymbol{k}_i$};
        \draw[->, >=latex, thick] (0, 0, 0) -- (0.75, 1, 0); \draw[-, thick] (0, 0, 0) -- (1.5, 2, 0); \node at (1.1, 0.8, 0) {$\boldsymbol{k}_o$};
        \def\r{0.2}
        \coordinate (A) at (\r,\r,\r);
        \coordinate (B) at (-\r,\r,\r);
        \coordinate (C) at (-\r,-\r,\r);
        \coordinate (D) at (\r,-\r,\r);

        \coordinate (E) at (\r,\r,-\r);
        \coordinate (F) at (-\r,\r,-\r);
        \coordinate (G) at (-\r,-\r,-\r);
        \coordinate (H) at (\r,-\r,-\r);

        \draw[thick] (A) -- (B) -- (C) -- (D) -- cycle;
        \draw[thick] (E) -- (F) -- (G) -- (H) -- cycle;
        \draw[thick] (A) -- (E);
        \draw[thick] (B) -- (F);
        \draw[thick] (C) -- (G);
        \draw[thick] (D) -- (H);

        \draw[dashed, thick] (0, 0, 0) -- (2, 0, 0);
        \node at (0.6, 0.3, 0) {$\alpha$};

        \draw[->, >=latex, thick] (-2.5, -0.5, 0) -- (-2.5, 0.5, 0); 
        \node at (-2.5, 0.7, 0) {$\hat{\boldsymbol{\varepsilon}}_{i\pi}$};
        \node at (-2.5, 0, 0) {$\boldsymbol{\odot}$};
        \node at (-2.9, 0, 0) {$\hat{\boldsymbol{\varepsilon}}_{i\sigma}$};

        \draw[->, >=latex, thick] (1.9, 1.7, 0) -- (1.1, 2.3, 0); 
        \node at (0.8, 2.3, 0) {$\hat{\boldsymbol{\varepsilon}}_{o\pi}$};
        \node at (1.5, 2, 0) {$\boldsymbol{\odot}$};
        \node at (1.9, 2.2, 0) {$\hat{\boldsymbol{\varepsilon}}_{o\sigma}$};

        \node at (-2, 2, 0) {$\boldsymbol{\odot}$};
        \node at (-2.2, 1.8, 0) {$z$};
        \draw[->, >=latex] (-2.5, 2, 0) -- (-1.5, 2, 0); \node at (-1.3, 2, 0) {$y$};
        \draw[->, >=latex] (-2, 1.5, 0) -- (-2, 2.5, 0); \node at (-2, 2.7, 0) {$x$};

        \draw (3.5, 0, 0) circle (1);
        \draw[->, >=latex, thick] (3.5, 0, 0) -- (3.5, 1, 0); \node at (3.5, 1.3, 0) {$\hat{\boldsymbol{\varepsilon}}_\pi$};
        \draw[->, >=latex, thick] (3.5, 0, 0) -- (4.5, 0, 0); \node at (4.8, 0, 0) {$\hat{\boldsymbol{\varepsilon}}_\sigma$};
        \draw[->, >=latex, thick] (3.5, 0, 0) -- (4.3, 0.6, 0); \node at (4.5, 0.8) {$\boldsymbol{\varepsilon}$};
        \node at (3.7, 0.5) {$\phi$};

        \draw (3.5, 0, 0) circle (0.11);
        \node at (3.5, 0, 0) {$\times$};
        \node at (3.2, 0, 0) {$\boldsymbol{k}$};
    \end{tikzpicture}
    \caption{Photon-in photon-out scattering geometry}
    \label{fig:AngleDefination}
\end{figure}

The relevant angles and directions are shown in Fig.~\ref{fig:AngleDefination}. The angle between the incident and scattered wavevectors, $\boldsymbol{k}_i$ and
$\boldsymbol{k}_o$, is $\alpha$, the scattering angle.  

For a fixed scattering
geometry, the incident and outgoing polarization vectors can be uniquely expressed
in terms of their $\sigma$ (perpendicular) and $\pi$ (parallel) components with
respect to the scattering plane spanned by $\boldsymbol{k}_i$ and $\boldsymbol{k}_o$:
\begin{equation}
\boldsymbol{\varepsilon}_{i,o} =  |\boldsymbol{\varepsilon}_{i,o}| \cdot (\hat{\boldsymbol{\varepsilon}}_{i,o}^\pi \cos{\phi_{i,o}} + \hat{\boldsymbol{\varepsilon}}_{i,o}^\sigma \sin{\phi_{i,o}})
\label{eq:IProj}
\end{equation}

\paragraph{E1E1 RIXS}

For linear polarization ($\boldsymbol{\varepsilon}_o^* = \boldsymbol{\varepsilon}_o$), Eq.~\ref{eq:sigmaDD} reduces to two components. By substituting  Eq.~\ref{eq:IProj}, the orientation average of the E1E1 RIXS intensity for linear polarizations is (up to a multiplicative factor):
\begin{equation}
    I_{\mathrm{E1E1}}(\alpha, \phi_i, \phi_o) = a + b(\sin{\phi_i}\sin{\phi_o}+\cos{\phi_i}\cos{\phi_o}\cos{\alpha})^2
   \label{E1E1_genI}
\end{equation}
where
\begin{eqnarray}
    a &=& \frac{1}{30} |\boldsymbol{\varepsilon}_i|^2|\boldsymbol{\varepsilon}_o|^2 \nonumber\\
    && (4\Tr(\mathbf{M}^\dagger\mathbf{M}) - |\Tr(\mathbf{M})|^2-\Tr (\mathbf{M}^*\mathbf{M})) \nonumber\\
    b &=& \frac{1}{30} |\boldsymbol{\varepsilon}_i|^2|\boldsymbol{\varepsilon}_o|^2 \nonumber\\&&(-2\Tr(\mathbf{M}^\dagger\mathbf{M}) + 3|\Tr(\mathbf{M})|^2+3\Tr
    (\mathbf{M}^*\mathbf{M})) 
    \nonumber \\
    \label{E1E1_ab}
\end{eqnarray}
By projecting on the polarization channels $\sigma$ ($\phi = \pi/2$) and $\pi$ ($\phi = 0$):
\begin{eqnarray}
    I_{\mathrm{E1E1}}^{\sigma\sigma} (\alpha) &=& a + b \nonumber\\
    I_{\mathrm{E1E1}}^{\sigma\pi} (\alpha) &=&  I_{\mathrm{E1E1}}^{\pi\sigma} (\alpha) = a \nonumber\\
    I_{\mathrm{E1E1}}^{\pi\pi} (\alpha)&=& a + b \cos^2{\alpha} 
    \label{E1E1_sp}
\end{eqnarray}
one can see the angular dependence is lost for $\sigma$-polarized incident light and for the crossed polarization channels.

When the outgoing polarization is not detected one must average over all $\phi_o$ values: 
\begin{eqnarray}
    I(\alpha, \phi_i) = \frac{1}{2\pi} \int_0^{2\pi} I(\alpha, \phi_i, \phi_o) \,d\phi_o 
\end{eqnarray}
In particular, from Eq.~\ref{E1E1_genI}:
\begin{eqnarray}
    I_{\mathrm{E1E1}}(\alpha,\phi_i) = a + \frac{1}{2} b (\sin^2{\phi_i} + \cos^2{\phi_i}\cos^2{\alpha}) \label{eq:IDD}
\end{eqnarray}
Subsequently, for $\sigma$ and $\pi$ incident polarizations:
\begin{eqnarray}
    I_{\mathrm{E1E1}}^\sigma (\alpha) &=& a + \frac{1}{2}b \nonumber\\
    I_{\mathrm{E1E1}}^\pi (\alpha) &=& a + \frac{1}{2}b \cos^2{\alpha}
\end{eqnarray}

\paragraph{E2E1 and E1E2 RIXS}

First, we address how many independent components are contained in the E2E1 and E1E2 terms for linear polarization. By reducing Eq. \ref{eq:sigmaDQ} according to $\boldsymbol{\varepsilon}_o^* = \boldsymbol{\varepsilon}_o$ we obtain:
\begin{eqnarray}
    &&\braket{\sigma_{\mathrm{RIXS}}^{\mathrm{E2E1}}} = \frac{1}{210} 
    \begin{bmatrix}
        |\boldsymbol{\varepsilon}_o|^2 |\boldsymbol{\varepsilon}_i|^2 |\boldsymbol{k}_i|^2 \\
        |\boldsymbol{\varepsilon}_o \cdot \boldsymbol{k}_i|^2 |\boldsymbol{\varepsilon}_i|^2 \\
        |\boldsymbol{\varepsilon}_o \cdot \boldsymbol{\varepsilon}_i|^2 |\boldsymbol{k}_i|^2 
    \end{bmatrix}^{\mathrm{T}} 
     \nonumber \\
    && \times 
    \begin{bmatrix}
        11 & -6 & -6 & -5 &   8 \\
        -6 &  9 &  9 &  4 & -12 \\
        -6 &  9 &  9 &  4 & -12 
    \end{bmatrix} \nonumber\\
    && \times\sum_{ii'jj'kk'}
    \begin{bmatrix}

        \delta_{ii'}\delta_{jj'}\delta_{kk'} \\ \delta_{ii'}\delta_{jk}\delta_{j'k'} \\
        \delta_{ii'}\delta_{jk'}\delta_{j'k} \\
        \delta_{ij}\delta_{i'j'}\delta_{kk'} \\ \delta_{ij}\delta_{i'k}\delta_{j'k'} \Re  
    \end{bmatrix}  S^*_{i'j'k'}S_{ijk}
   \label{eq:sigmaDQ_linear}
\end{eqnarray}
Although Eq. \ref{eq:sigmaDQ_linear} is written with three rows - the minimal dimension of the bilinear form - two of them correspond to the same linear combination of $\delta$-functions. Consequently, only two components are truly independent.

By plugging Eq.~\ref{eq:IProj} into Eq.~\ref{eq:sigmaDQ_linear}, the orientation averages of the E2E1 and E1E2 RIXS read:
\begin{eqnarray}
    I_{\mathrm{E2E1}}(\alpha, \phi_i, \phi_o) &=& a + b\cos^2{\phi_o}\sin^2{\alpha} \nonumber\\ &+& b(\sin{\phi_i}\sin{\phi_o}+  \cos{\phi_i}\cos{\phi_o}\cos{\alpha})^2 \nonumber
\end{eqnarray}
\begin{eqnarray}
    I_{\mathrm{E1E2}}(\alpha, \phi_i, \phi_o) &=& a + b\cos^2{\phi_i}\sin^2{\alpha} \\
    & + & b(\sin{\phi_i}\sin{\phi_o}+\cos{\phi_i}\cos{\phi_o}\cos{\alpha})^2 \nonumber
\end{eqnarray}
where $a$ and $b$ can be inferred from Eq.~\ref{eq:sigmaDQ_linear}. 

By projecting on the individual channels:
\begin{eqnarray}
    I_{\mathrm{E2E1}}^{\sigma\sigma} &=& I_{\mathrm{E2E1}}^{\pi\pi} =
    I_{\mathrm{E1E2}}^{\sigma\sigma} = I_{\mathrm{E1E2}}^{\pi\pi} =
    a + b \nonumber\\
    I_{\mathrm{E2E1}}^{\sigma\pi}&=& I_{\mathrm{E1E2}}^{\pi\sigma} = a + b \sin^2{\alpha} \nonumber\\
    I_{\mathrm{E2E1}}^{\pi\sigma} &=& I_{\mathrm{E1E2}}^{\sigma\pi} = a 
\end{eqnarray}
When the outgoing polarization is not detected:
\begin{eqnarray}
    I_{\mathrm{E2E1}}(\alpha, \phi_i) = a + \frac{b}{2} ( 1 +  \sin^2{\phi_i}\sin^2{\alpha}) \label{eq:IQD}\nonumber \\
    I_{\mathrm{E1E2}}(\alpha, \phi_i)  = a + \frac{b}{2} ( 1 + \cos^2{\phi_i}\sin^2{\alpha}) \label{eq:IDQ}
\end{eqnarray}
In particular, for $\sigma$ and $\pi$ incident polarizations:
\begin{eqnarray}
    I_{\mathrm{E2E1}}^\sigma(\alpha) &=& I_{\mathrm{E1E2}}^\pi(\alpha) = a + \frac{b}{2} (1 + \sin^2{\alpha}) \nonumber\\
    I_{\mathrm{E2E1}}^\pi(\alpha) &=& I_{\mathrm{E1E2}}^\sigma(\alpha) = a + \frac{b}{2}  
    \label{eq:IDQ_sp}
\end{eqnarray}

\subsubsection{Scattering from spherically symmetric states}

Consider the non-resonant, elastic scattering cross section $\sigma_{\mathrm{Th}}$, i.e. the Thomson term. For linearly polarized light:
\begin{equation}
    \sigma_{\mathrm{Th}}(q) =  r_0^2 |f_0(q)|^2 |\hat{\boldsymbol{\varepsilon}}_o\cdot\hat{\boldsymbol{\varepsilon}}_i|^2 \label{eq:Thomson}
\end{equation}
where $r_0$ is the classical electron radius, $q$ the momentum transfer  and $f_0(q)$ the atomic form factor. Thomson scattering originates from the spherically symmetric part of the electronic charge density and is therefore rotational invariant. Similarly, when the electron states $f$ and $g$ are spherically symmetric we retrieve a special case of Eq.~\ref{eq:sigmaDD} where the scattering tensor is:
\begin{equation}
    \mathbf{M} \propto \begin{bmatrix}
        1 & 0 & 0 \\
        0 & 1 & 0 \\
        0 & 0 & 1
    \end{bmatrix}
\end{equation}
This will give a coefficient $a=0$ in Eq.~\ref{E1E1_ab} with which the angular dependence in Eq.~\ref{E1E1_genI} becomes Thomson-like, i.e. similar to Eq.~\ref{eq:Thomson}. Furthermore, like in the Thomson scattering, the crossed polarization $\sigma\pi$ and $\pi\sigma$ terms in Eq.~\ref{E1E1_sp} are forbidden.

Here we have treated the resonant scalar scattering and Thomson scattering on equal footing, as they have identical tensor structures. Nevertheless the classical diffraction limit cannot rigorously be retrieved from our theory, as they stem from different Hamiltonian terms \cite{Sakurai_Advanced_1967}. 

\section{Application}

We apply our method to study the angular dependence of two types of RIXS processes at the \ce{L3} edge of \ce{CeO2}. The first is core-to-core (ctc) 2$p$3$d$ RIXS, in particular L$_3$M$_5$ RIXS (the L$_{\alpha 1}$ line, i.e. the $3d_{5/2} \rightarrow 2p_{3/2}$ resonant emission) and L$_3$M$_4$ RIXS (the L$_{\alpha 2}$ line, i.e. $3d_{3/2} \rightarrow 2p_{3/2}$). The second is valence-to-core (vtc) RIXS, corresponding to valence $ \rightarrow 2p_{3/2}$ transitions.

\subsection{Methods}

\subsubsection{Experiment setup}
\begin{figure}[!htbp]
\centering
\begin{tikzpicture}
    \fill[rotate around={-45:(0,0)}, gray] (-0.5,-0.05) rectangle (0.5,0.05);
    \draw[->, >=latex, thick] (0,2.0) -- (0,0); \node at (-0.3,1.2) {$\boldsymbol{k}_i$};
    \draw[->, >=latex, thick] (0,0) -- (2.5,0); \node at (1.5,-0.3) {$\boldsymbol{k}_o$};
    \draw[dashed, thick] (0,0) -- (0,-1); \node at (0.25,-0.5) {$\alpha$};
    \draw[thick] (0.25,0) arc[start angle=0, end angle=-90, radius=0.25];
    \draw[->, >=latex, thick] (-0.5,2) -- (0.5,2); \node at (0.3,2.3) {$\boldsymbol{\varepsilon}_i$};

    \fill[rotate around={90:(3,0)}, blue] (3,0) ellipse (0.2 and 0.05);
    \node[anchor=west] at (3.35,0) {$90^\circ$};

    \fill[rotate around={100:(2.954,0.521)}, blue] (2.954,0.521) ellipse (0.2 and 0.05);
    \node[anchor=west] at (3.25,0.57) {$97^\circ$};

    \fill[rotate around={80:(2.954,-0.521)}, blue] (2.954,-0.521) ellipse (0.2 and 0.05);
    \node[anchor=west] at (3.25,-0.57) {$83^\circ$};

    \fill[rotate around={110:(2.819,1.026)}, blue] (2.819,1.026) ellipse (0.2 and 0.05);
    \node[anchor=west] at (3.10,1.08) {$104^\circ$};

    \fill[rotate around={70:(2.819,-1.026)}, blue] (2.819,-1.026) ellipse (0.2 and 0.05);
    \node[anchor=west] at (3.10,-1.08) {$76^\circ$};

    \fill[rotate around={90:(3.1,0)}, red] (3.1,0) ellipse (0.2 and 0.05);
    \fill[rotate around={100:(3.053,0.538)}, red] (3.053,0.538) ellipse (0.2 and 0.05);
    \fill[rotate around={110:(2.912,1.061)}, red] (2.912,1.061) ellipse (0.2 and 0.05);

    \fill[rotate around={120:(2.683,1.550)}, red] (2.683,1.550) ellipse (0.2 and 0.05);
    \node[anchor=west] at (2.95,1.60) {$111^\circ$};

    \fill[rotate around={130:(2.376,1.995)}, red] (2.376,1.995) ellipse (0.2 and 0.05);
    \node[anchor=west] at (2.65,2.05) {$118^\circ$};
\end{tikzpicture}
    \caption{Positioning of the analyzer crystals with respect to the sample (blue for the vtc measurement, red for ctc).}
    \label{fig:analyzers}
\end{figure}
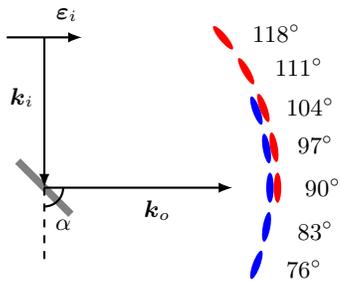

The measurements were performed at beamline ID26 of the European Synchrotron Research Facility, using a spectrometer set in a vertical Rowland geometry \cite{Glatzel_The_2021} with sample, detector and five crystal analyzers arranged on 1m radius circles. The round analyzer crystals have a diameter of 100 mm. The incoming X-ray polarization is horizontal, parallel to the scattering plane ($\pi$ incident polarization) and no polarization analysis was performed. Spectra were collected by
displacing a pixel detector (Dectris Pilatus 100k) by 40~mm from the nominal focusing condition outside the Rowland circle. In this configuration, each analyzer produces a distinct spot on the detector allowing 5 scattering angles to be recorded simultaneously (Fig.~\ref{fig:analyzers}).

The incident energy was scanned using a Si(311) double crystal monochromator. The beam size was 0.2 mm (horizontal) x 0.1 (vertical) and the sample surface was oriented at approximately 45 degrees relative to the incoming beam. The rotation of the (powder) sample does not affect the angular distribution of the intensity, beyond self-absorption effects. RIXS planes were acquired by continuously scanning the monochromator energy for each analyzer position, while the undulator gap was tuned to maximize the incident photon flux on the first harmonic. Higher harmonics were rejected by total reflection on Si mirrors inclined at 2.5 mrad with respect to the incident beam. The incident intensity used for normalization was monitored in real time with a photo diode operating in backscattering geometry. One-dimensional cuts of the RIXS planes
were obtained by binning the data directly, without applying any smoothing or interpolation procedures \cite{DAXS}.

The Ge(333) reflection at 84$^\circ$ Bragg angle was used for the vtc RIXS
measurements, yielding an overall energy resolution of 0.5~eV, defined as the full
width at half maximum of the elastic line. The analyzer crystals were arranged in a symmetric configuration, with the central
analyzer positioned at a scattering angle of $90^\circ$ and the remaining analyzers
placed at $90^\circ \pm 7^\circ$ and $90^\circ \pm 14^\circ$ (Fig.~\ref{fig:analyzers}). Note that the diameter of an analyzer crystal covers an angle of $\approx6.4^\circ$. For measurements of the 2$p$3$d$ RIXS the Ge(331) reflection at 81$^\circ$ was used, providing a total energy resolution of
0.8~eV. In this case, the spectrometer was rotated so that the analyzers were set to cover a 28$^\circ$ range, from 90$^\circ$ to 118$^\circ$ (Fig.~\ref{fig:analyzers}). The incoming energy was calibrated by measuring the K absorption edge of a Ti foil.  The energy calibration of the scattered X-rays is based on the geometric alignment of the spectrometer components. In the case of vtc RIXS, the elastic scattering is used to achieve absolute calibration of the energy transfer.

The CeO$_2$ powder was purchased commercially and pressed to a pellet. No radiation damage was noticed during the experiment.

\subsubsection{Computational method}

Transition tensors were calculated by OpenMolcas v25.06 \cite{Fdez_OpenMolcas_2019, Aquilante_Modern_2020}. The embedded cluster model of \ce{CeO2} was constructed based on crystal structures of ICSD 88759 \cite{Kummerle_The_1999}. Zone 1 (full-basis atoms) consisted of a \ce{Ce}-centered \ce{CeO_8^{12-}} cluster. Zone 2 contained 400 fixed point charges, while Zone 3 contained 1000 variable point charges used to balance the electrostatic potential of Zones 1 and 2. The point charges were generated using the Ewald program \cite{Derenzo_Determining_2000, Klintenberg_Accurate_2000}, based on a search within a $\mathrm{12 \times 12 \times 12}$ supercell, assigning formal charges of $+4$ to \ce{Ce^{4+}}, and $-2$ to \ce{O^{2-}}. In the subsequent calculations, point charges located within 3 \AA\ of the cluster were replaced by effective core potentials \cite{Pascual_Ab_1993, Nygren_Bonding_1994}, as described in Ref.\cite{Gendron_Puzzling_2017}.

Restricted/complete active space self-consistent field (RASSCF/CASSCF) approaches \cite{Roos_The_1987, Roos_A_1980, Siegbahn_A_1980, Siegbahn_The_1981, Malmqvist_The_1990} were employed to obtain the wavefunctions of different electronic states. Relativistic effects were incorporated using the second-order Douglas–Kroll–Hess (DKH2) Hamiltonian \cite{Douglas_Quantum_1974, Hess_Applicability_1985, Hess_Relativistic_1986, Wolf_The_2002}, in combination with all-electron atomic natural orbital relativistic core-correlated basis sets of valence triple-$\zeta$ quality (ANO-RCC-VTZP). The resolution-of-the-identity Cholesky decomposition technique \cite{Pedersen_Density_2009} was applied to accelerate the evaluation of two-electron integrals. A finite nuclear model \cite{Dyall_Finite_1993} was also employed. Before the transition calculation, the dynamic correlation is corrected by single state second-order perturbation approach (SS-CASPT2) \cite{Andersson_Second_1992}. Transitions were then calculated using the Restricted Active Space State Interaction (RASSI) method with spin–orbit coupling (SOC) \cite{Malmqvist_The_2002}. Both singlet and triplet states were included in the SOC-RASSI calculations. 

The in-house program Polarixs, available as Python package \cite{Polarixs}, was used to convolute the matrix elements extracted from OpenMolcas to obtain the tensor in Eq.~\ref{eq:defM} and subsequently estimate the first and second order polarization dependent cross sections by Eq.~\ref{eq:E_conv}. The broadening effects of $\Gamma_n$ (width of the intermediate states in Eq.~\ref{eq:defM}) and $\Gamma_f$ (width of the final states in Eqs.~\ref{eq:Fermi} - \ref{eq:E_conv}) are implemented numerically by convolution with Lorentzian functions of width 2$\Gamma_n$ and 2$\Gamma_f$ respectively. 

The energy scale of the calculation was downshifted by $32$ eV to align with the experiment.

\subsection{Results and discussion}

\ce{CeO2} crystallizes in the $Fm\bar{3}m$ space group, with Ce ions occupying
sites of $\mathrm{O_h}$ point symmetry. The Ce 4$f$ states hybridize with O 2$p$
orbitals to form a bonding state, corresponding to the
ground state $\lvert g\rangle$, and an antibonding state \cite{Sergentu_Probing_2021, Sergentu_Xray_2022}. Due to
the closed-shell configuration, both the ground state $\lvert g\rangle$ and the
antibonding state are totally symmetric and thus transform according to the
$A_{1g}$ irreducible representation.  

\subsubsection{Core-to-core L$_3$M$_5$ RIXS}

\begin{figure}[!htbp]
    \centering
    \includegraphics[width=0.9\linewidth]{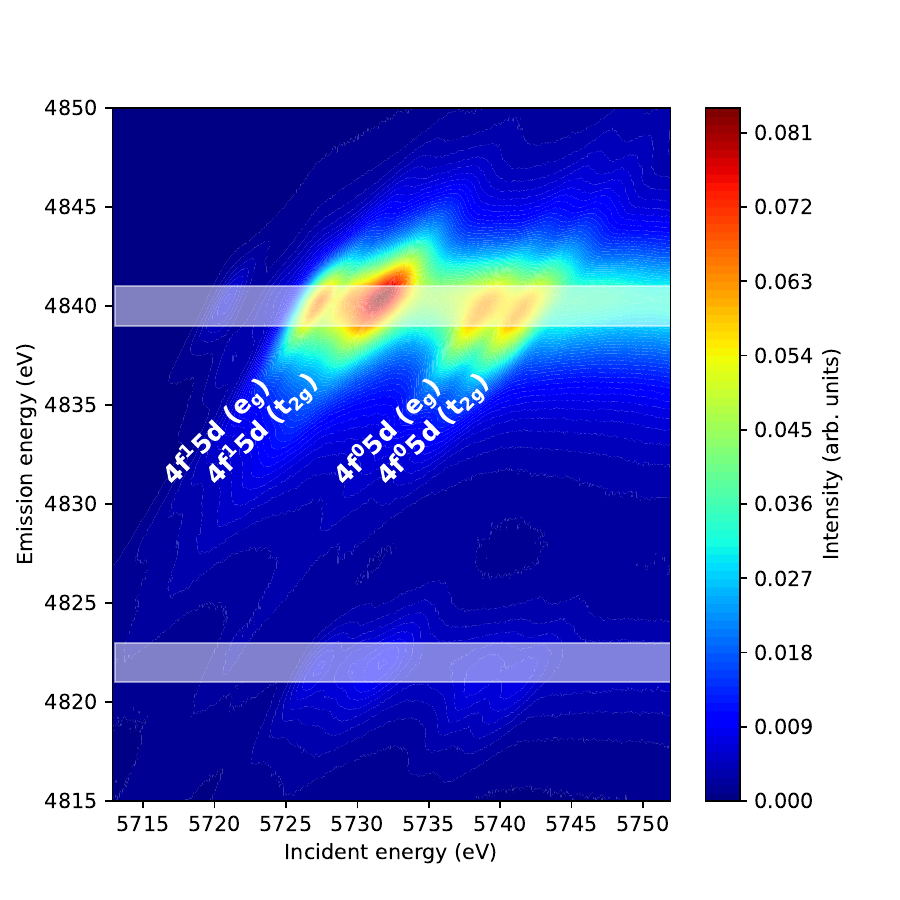}
    \caption{Experimental 2$p$3$d$ RIXS for the $\alpha = 118^\circ$ analyzer. Features at 4840 and 4822 eV emission energies correspond to the L$_3$M$_5$ and  L$_3$M$_4$ lines respectively. Peaks are annotated according to the final state they are originating from. The displayed masks were used to obtain the CEE cuts in Fig.~\ref{fig:L3M45_cut}.}
    \label{fig:2p3dRIXS}
\end{figure}

In Fig.~\ref{fig:2p3dRIXS} we show the ctc RIXS plane for one of the analyzers. The main features at 4840 eV emission energy correspond to the L$_3$M$_5$ resonant emission line ($3d_{5/2}\rightarrow2p_{3/2}$), while the weaker features at 4822 eV are the L$_3$M$_4$ (L$_{\alpha2}$) line ($3d_{3/2}\rightarrow2p_{3/2}$).  The transition paths for the L$_3$M$_5$ RIXS process are $$\ket{g}\to\ket{2p^54f^{0/1}5d^1}\to\ket{3d^94f^{0/1}5d^1}$$ and were discussed in detail in Ref. \cite{Kotani_Spectator_2012}.  Subsequently, the main
features in the RIXS plane arise from E1E1 processes and can be grouped into two sets,
corresponding to final-state configurations with $\mathrm{4f^1}$ and $\mathrm{4f^0}$ character,
respectively (see annotations for L$_3$M$_5$ in Fig.~\ref{fig:2p3dRIXS}). Within each group, the crystal-field splitting between the $e_g$
and $t_{2g}$ components is clearly resolved.

\begin{figure}[!htbp]
    \centering
        \includegraphics[width=0.9\linewidth]{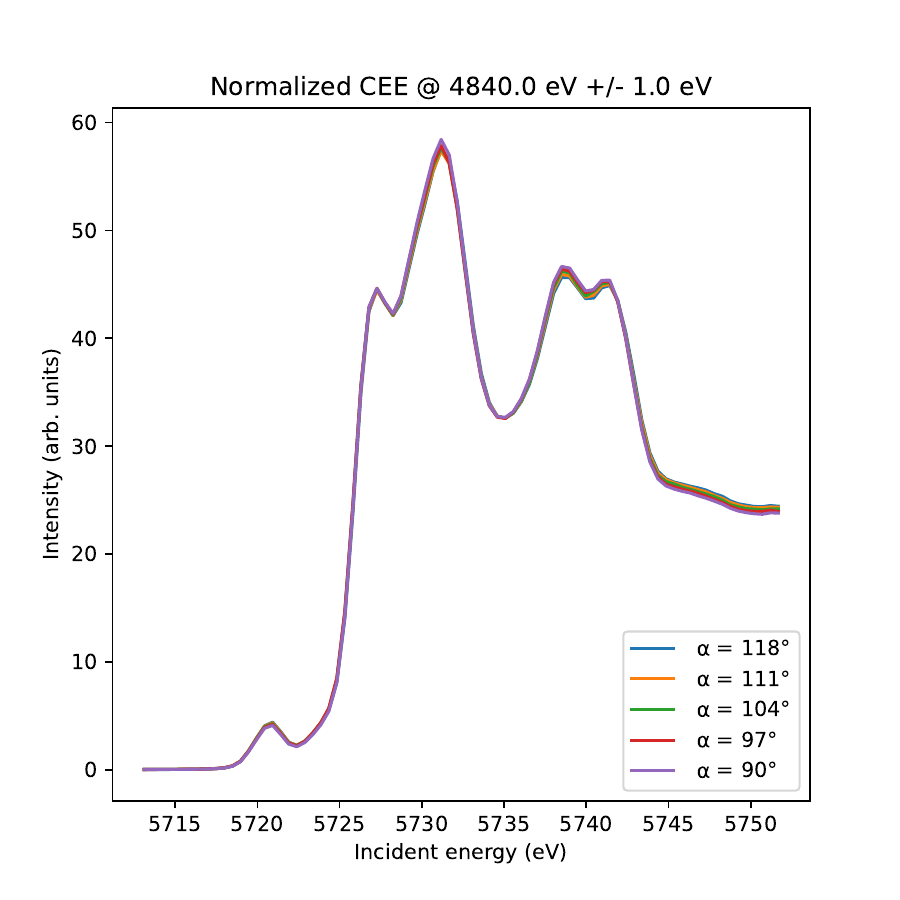}
        \includegraphics[width=0.9\linewidth]{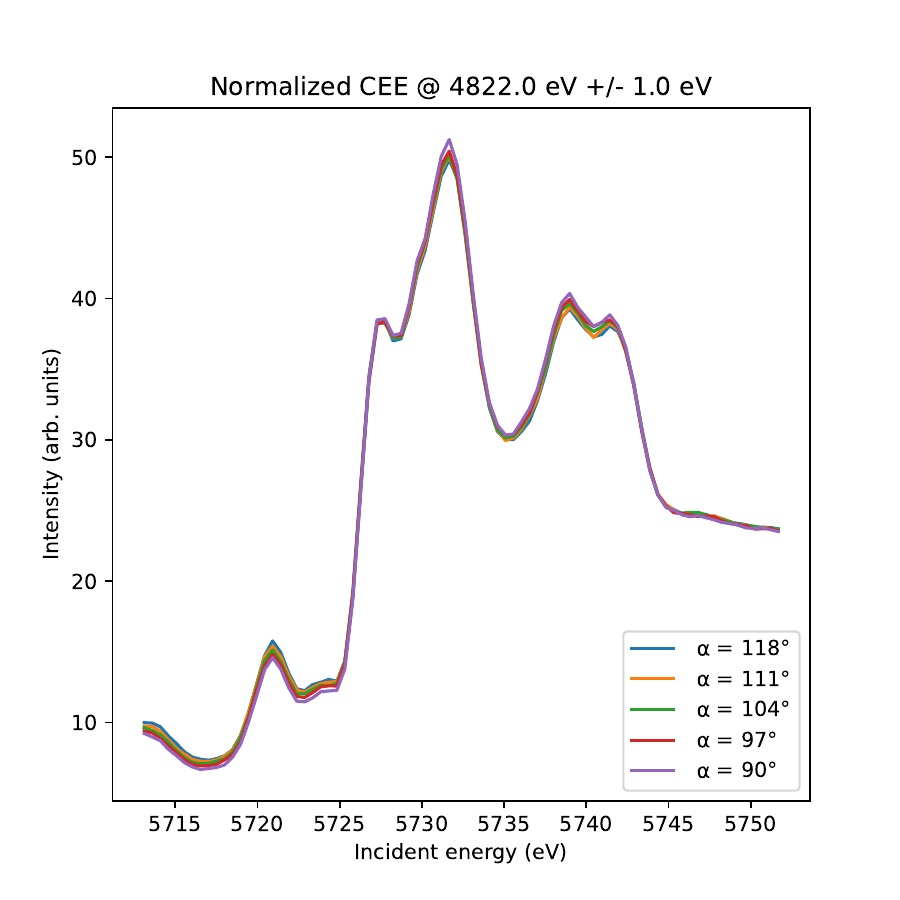}
    \caption{CEE cuts for all analyzers obtained with the masks in Fig.~\ref{fig:2p3dRIXS}.  There is no significant angular dependence for L$_3$M$_5$ (top)  and L$_3$M$_4$ RIXS (bottom).}
    \label{fig:L3M45_cut}
\end{figure}

The constant emission energy (CEE) cuts are shown in Fig.~\ref{fig:L3M45_cut}. All cuts were normalized to their respective area, which is intended to compensate for non-identical reflection intensity of the analyzer crystals and partially correct for self-absorption effects that depend on the scattering angle. For both emission lines, the RIXS intensities do not exhibit a genuine dependence on the scattering angle $\alpha$.  

Our OpenMolcas analysis indicates the pre-edge feature originates from an E2E1
process ($\mathrm{2p} \to \mathrm{4f}$). For the L$_3$M$_5$ cuts (Fig.~\ref{fig:L3M45_cut}, top), the pre-edge intensity is strictly constant as a function of the scattering angle.  When the absorbing atom sits on an inversion center, intraatomic $p$-$d$ hybridization is symmetry forbidden and therefore pre-edge peaks are essentially E2E1. For $\pi$ incoming polarization,  Eq.~\ref{eq:IDQ_sp} predicts a pre-peak intensity independent on the scattering angle, which is consistent with the L$_3$M$_5$ experimental data. On the other hand, one can see from Fig.~\ref{fig:2p3dRIXS} that the pre-edge structure of L$_3$M$_4$ is affected by the tails of the E1E1 peaks above, belonging to L$_3$M$_5$. This explains why the L$_3$M$_4$ pre-edge structure is more prominent than the L$_3$M$_5$ one, and its subsequent slight angular dependence.

\begin{figure*}[t]
    \centering
    \includegraphics[width=\textwidth]{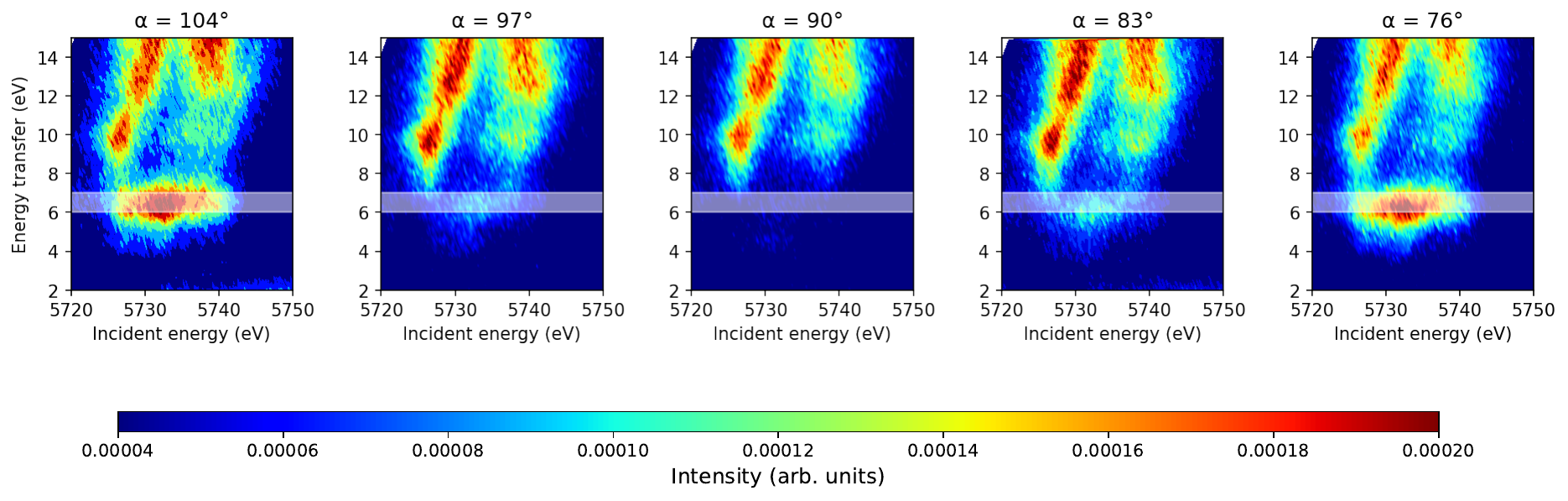}
    \caption{Valence-to-core RIXS planes recorded for each analyzer are shown. The applied mask isolates the angular-dependent feature at an energy loss of 6.5 eV, with a bandwidth of 1 eV.}
    \label{fig:vtcRIXS}
\end{figure*}

\begin{figure}[!htbp]
    \centering
    \includegraphics[width=0.9\linewidth]{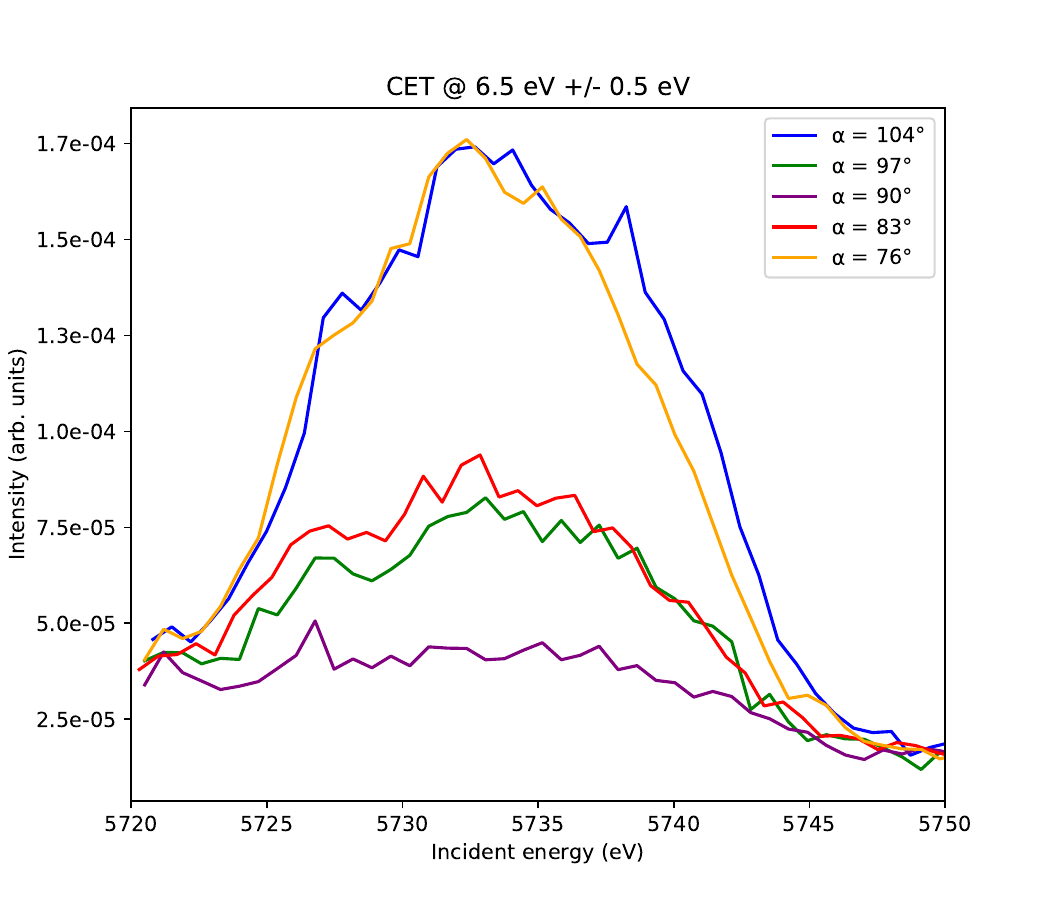}
    \includegraphics[width=0.9\linewidth]{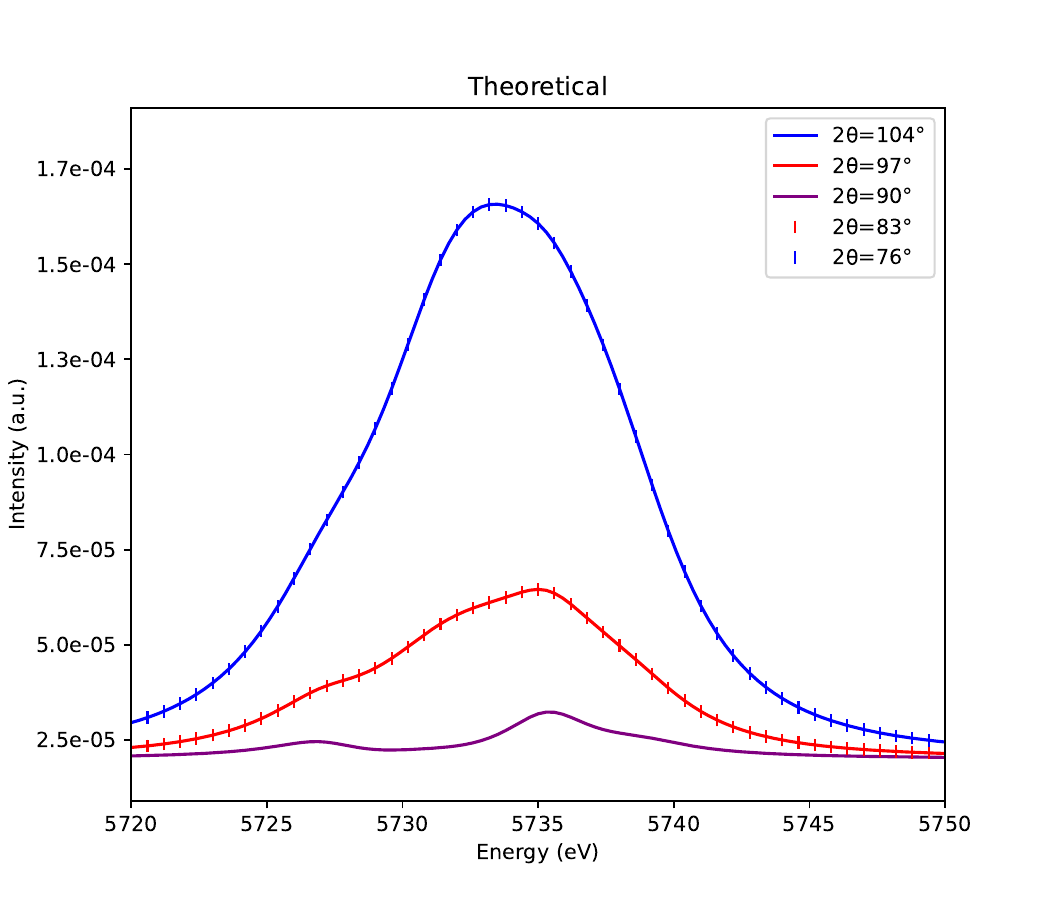}
    \caption{Angular dependence in vtc RIXS. Top: constant energy transfer cuts in the measured vtc RIXS planes with the mask displayed in Fig.~\ref{fig:vtcRIXS}. Those analyzers disposed symmetrically wrt. 90$^\circ$ have similar signals. Bottom: angular dependence predicted from calculations. The signals with $\alpha = 90^\circ + \Delta\alpha$ and $\alpha = 90^\circ - \Delta\alpha$ are strictly identical.}
    \label{fig:vtcRIXScut}
\end{figure}
While the CEE cuts in Fig.~\ref{fig:2p3dRIXS} are consistent with the result in Eq.~\ref{eq:IDQ_sp}, namely that E2E1 features in $\pi$ incident polarizations are independent of the scattering angle, the absence of an angular dependence in the E1E1 peaks is less immediately apparent.
Although we unambiguously reproduce this
behavior numerically (not shown), we further support these results with an analytical explanation.
Such approach is tractable in high-symmetry situations, e.g. in  $O_h$ symmetry, but generally becomes impractical for lower-symmetry
environments or for systems with higher electronic occupation numbers. The derivation is discussed in appendices~\ref{sec:AppendixA} and~\ref{sec:AppendixB}.
For both transition paths we obtain:
\begin{eqnarray}
    I_{3d^95d_{e_g}^1} &=& \frac{|\boldsymbol{\varepsilon}_i|^2|\boldsymbol{\varepsilon}_o|^2}{30}\times \frac{66+\cos^2{\alpha}}{45} R_{2p3d}^2 \nonumber\\
    I_{3d^95d_{t_{2g}}^1} &=& \frac{|\boldsymbol{\varepsilon}_i|^2|\boldsymbol{\varepsilon}_o|^2}{30}\times \frac{66+\cos^2{\alpha}}{30} R_{2p3d}^2
    \label{eq:Ictc}
\end{eqnarray}
These results clearly show that, even if the $\alpha$ dependent term is non-zero, its contribution is much weaker than the constant term and therefore the angular dependence is not seen in the experiment.

\subsubsection{Valence-to-core RIXS}

The measured vtc RIXS planes are shown in Fig.~\ref{fig:vtcRIXS}. The corresponding theoretical planes were computed by convoluting the transition amplitudes obtained with OpenMolcas for the different analyzer orientations. Reasonable agreement is obtained; however, a detailed discussion is deferred to future work, as the present study focuses exclusively on the angular dependence.

The only RIXS feature exhibiting obvious angular dependence is situated at 6.5 eV energy loss (Fig.~\ref{fig:vtcRIXS}). The corresponding constant energy transfer (CET) cuts are shown in Fig.~\ref{fig:vtcRIXScut} (top). Our calculations quantitatively reproduce this result (Fig.~\ref{fig:vtcRIXScut}, bottom).  

The transition path associated to this peak involves intermediate states of type $2p^54f^{0/1}5d^1$, all decaying to the antibonding state. Alternatively, one can describe this process as the excitation of a $2p$ electron to a $d$ state, followed by its decay that fills the core hole and, due to the core-hole interaction, concomitantly triggers a valence band excitation (i.e. an electron is promoted from the bonding to the antibonding state). For this pathway, the same electron is involved in both absorption and emission steps of the RIXS process - in other words, the final state contains no $5d$ excitation. Depending on the authors, this is referred as a \emph{participator} \cite{Kotani_Spectator_2012} or \emph{indirect} RIXS process  \cite{Brink_Theory_2005, Ament_Resonant_2011}, in contrast to \emph{spectator}
 or \emph{direct} RIXS processes, which involve two distinct electrons.  

From a group-theoretical perspective, the transition pathway associated to the 6.5~eV energy loss peak is peculiar in that it connects initial and final states of $A_{1g}$ symmetry. In this case the scattering is isotropic (Thomson-like) with the following structure of the $\mathbf{M}$ tensor:
\begin{equation}
    M_{A_{1g}} = \frac{1}{\sqrt{3}}\begin{bmatrix}
        1 & 0 & 0 \\
        0 & 1 & 0 \\
        0 & 0 & 1
    \end{bmatrix} \nonumber
\end{equation}
and the angular dependence:
\begin{equation}
    I_{A_{1g}}(\alpha) = \frac{|\boldsymbol{\varepsilon}_i|^2|\boldsymbol{\varepsilon}_o|^2}{30}\times \frac{1}{6}R_{A_{1g}}^2 \cos^2{\alpha} 
    \label{eq:I_antibonding}
\end{equation}
where $R_{A_{1g}}$ is a coefficient proper to this particular final state. For this peak, the angular dependence is the most pronounced, as the tensor structure enforces a cancellation of the constant term ($a = 0$ in Eq.~\ref{E1E1_ab}). 

The angular dependence given in Eq.~\ref{eq:I_antibonding} predicts a complete
suppression of the intensity for the central analyzer at
$\alpha = 90^\circ$. While this behavior is qualitatively observed in the
measured RIXS planes (Fig.~\ref{fig:vtcRIXS}), the corresponding cut shown in
Fig.~\ref{fig:vtcRIXScut} exhibits a small residual intensity. This background
signal originates from the finite size of the analyzer crystals, on one hand, as well as from the tails of neighboring spectral features, on the other hand. The latter can be reproduced by the Polarixs calculations (see Fig.~\ref{fig:vtcRIXScut}, bottom). We deliberately selected convolution widths ($\Gamma_n = 4$ eV for the intermediate states, $\Gamma_f = 1.5$ eV for the final states) that reveal the fine structure in the CET cuts. Increasing $\Gamma_n$ would progressively smear out these features. A similar analysis establishing the lack of angular dependence of the direct RIXS peaks is provided in the Supplementary Information.

The application of our framework to vtc RIXS at the Ce L$_3$ edge in
\ce{CeO2} is consistent with earlier results reported by Kotani et al. \cite{Kotani_Spectator_2012}. They were first to accurately explain the origin of the L$_3$M$_5$ and vtc RIXS features, which are confirmed by our current study with OpenMolcas. To explain the pronounced angular dependence of participator processes - absent in spectator processes -, Kotani et al. employed simplified
concepts based on a spherical-tensor decomposition (Eq.~7 of
Ref.~\cite{Kotani_Spectator_2012}), combined with the angular dependence derived in
an earlier work (Eq.~2.13 of Ref.~\cite{Nakazawa_Theory_2000}). Within this framework,
participator processes were argued to selectively probe a specific,
angular-dependent component of the spherical tensor, whereas for spectator
processes no such selection occurs, leading to a cancellation of the angular
dependence. While Ref.~\cite{Kotani_Spectator_2012} successfully accounts for the experimental
observations, this early study suffers from several limitations. In particular,
the scattering process is treated within an isotropic approximation, which is adequate for the 6.5 eV energy loss transition owing to the
$A_{1g}$ character of the corresponding final state, but fails to capture the
tensor nature of the scattering for more general final state symmetries. From a
group-theoretical perspective, the angular expression employed is incomplete and
therefore cannot describe the polarization dependence in more complex situations.
Specifically, an expansion up to second order would have been required in Eq.~7 of
Ref.~\cite{Kotani_Spectator_2012} to render the formulation more general; even then,
its applicability would remain restricted to systems with $O_h$ symmetry.

\section{Conclusion}

This work presents a practical approach to interpreting angular and polarization dependent XAS and RIXS spectra in orientation-averaged systems, such as liquid or powder samples. By working directly with Cartesian transition tensors—the natural output of modern quantum chemistry codes—our approach bypasses the complexity of spherical tensor formalisms while maintaining full generality and rigor. 

The key advantages of this framework are threefold. First, it provides explicit, closed-form expressions for orientation-averaged intensities that cleanly separate experimental geometry from sample electronic structure, enabling straightforward interpretation of polarization-dependent measurements. Second, it handles arbitrary point-group symmetries without requiring case-by-case derivations, making it broadly applicable across diverse materials systems. Third, it seamlessly integrates with ab initio calculations from packages like OpenMolcas and ORCA, eliminating the need for coordinate transformations and enabling quantitative predictions of spectral features.

We present the first derivation of RIXS powder averages within a Cartesian tensor framework, providing explicit formulae for the E1E1, E1E2, and E2E1 contributions, together with a systematic methodology for obtaining the E2E2 term and even higher order terms. We project these averages onto the fundamental linear polarizations $\sigma$ and $\pi$, and thereby extract the principal components (two for both E1E1 and E1E2/E2E1).  

We validated our approach through detailed analysis of RIXS measurements at the Ce L$_3$ edge in \ce{CeO2}. The framework successfully reproduces the absence of angular dependence in L$_3$M$_5$ RIXS and explains the pronounced $\cos^2{\alpha}$ behavior of the 6.5 eV vtc feature, arising from the ground state and final state sharing the same $A_{1g}$ symmetry.
For verification and educational purposes, group theory was employed to deduce the analytical angular dependence for each transition.

Our formalism naturally treats circular polarization for electric multipole transitions (E1 and E2) and reproduces the known result that E1E2 interference in absorption vanishes upon orientational averaging.

While the current formulation addresses non-magnetic and non-chiral systems, the theoretical framework introduced in this work is already capable of addressing certain cases beyond these limits, such as magnetic scattering without additional transition operators and single-handed chiral structures (since Eq.~\ref{eq:SO3int} assumes the crystal is generated by SO(3) operators). The tensor-based foundation naturally extends to more complex scenarios. Future work incorporating complete magnetic and chiral contributions will require including the magnetic terms in the transition operator, nonetheless the Cartesian framework established here provides a clear pathway forward. More generally, this methodology applies to a wide range of X-ray spectroscopies - including non-resonant X-ray Raman scattering - as well as to optical spectroscopies such as Raman and UV–Vis. It can be straightforwardly extended to higher-order multipole processes (E2E2, E3E1, etc.) within the same systematic framework.

\begin{acknowledgments}
We thank Blanka Detlefs for help with the data acquisition and processing and Vasyl Marchuk for preparing the sample. One of the authors (O. Bun\u au) thanks Yves Joly and Andrei Rogalev for fruitful discussions. This work was co-financed by the European Union through an Erasmus Mundus grant from the Education, Audiovisual, and Culture Executive Agency (EACEA).
\end{acknowledgments}

\appendix

\section{Polarization dependence of different irreducible representations under $O_h$ symmetry} \label{sec:AppendixA}

A symmetry analysis of the tensor elements defined in Eq.~\ref{eq:defM} shows that when the ground state $\ket{g}$ belongs to $A_{1g}$, both the E1E1
scattering tensor and the accessible final states $\lvert f\rangle$ transform
according to the direct product
\begin{equation}
    T_{1u} \otimes T_{1u} = A_{1g} \oplus E_g \oplus T_{1g} \oplus T_{2g}
\end{equation}

In particular, the $\mathbf{M}$ tensor structure is fully defined by the irreducible representation (irrep) the corresponding final state belongs to. 

Every peak in the RIXS plane generally contains the summed contributions from various final states. Because the summation over final states appears outside the modulus squared in
Eq.~\ref{eq:KramersHeisenberg}, final states do not mix with each other. 
Therefore, the polarization dependence must be analyzed separately for each final state and then summed up together. Depending on the decomposition of the final state, not all irreps are necessarily present in the sum, or they may be represented more than once.

 This section gives (a) the angular dependence of each irrep (for simplicity, as $\pi$-incident without polarization detection) and (b) the basis matrices for the $\mathbf{M}$ tensor description.

If the final state is $A_{1g}$, the transition tensor is isotropic and described by the diagonal matrix:
\begin{equation}
    \frac{1}{\sqrt{3}}\begin{bmatrix}
        1 & 0 & 0 \\
        0 & 1 & 0 \\
        0 & 0 & 1
    \end{bmatrix} \nonumber
\end{equation}
and the angular dependence:
\begin{equation}
    I_{A_{1g}} = \frac{|\boldsymbol{\varepsilon}_i|^2|\boldsymbol{\varepsilon}_o|^2}{30}\times \frac{1}{6}R_{A_{1g}}^2 \cos^2{\alpha} 
\end{equation}

Here, $R_{A_{1g}}$ is a coefficient depending on the final state and its irrep.

The transition to the $E_g$ irrep spans a twofold-degenerate subspace, for
which two commonly used basis tensors are
\begin{equation}
    \frac{1}{\sqrt{2}}\begin{bmatrix}
        1 & 0 & 0 \\
        0 & -1 & 0 \\
        0 & 0 &  0
    \end{bmatrix}\quad \text{and} \quad
    \frac{1}{\sqrt{6}}\begin{bmatrix}
        -1 & 0 & 0 \\
        0 & -1 & 0 \\
        0 & 0 &  2
    \end{bmatrix} \nonumber
\end{equation}

A noteworthy property of the $E_g$ representation is that any choice of basis within
this subspace, leads to the same polarization dependence, namely:
\begin{equation}
    I_{E_{g}} = \frac{|\boldsymbol{\varepsilon}_i|^2|\boldsymbol{\varepsilon}_o|^2}{30}\times \frac{1}{10}R_{E_g}^2\left(1 + \frac{1}{6} \cos^2{\alpha}\right)
\end{equation}

Similarly, $T_{2g}$ is three-fold degenerate and can only give off-symmetric diagonal matrices like:
\begin{equation}
    \frac{1}{\sqrt{2}}\begin{bmatrix}
        0 & 0 & 0 \\
        0 & 0 & 1 \\
        0 & 1 & 0
    \end{bmatrix} \ \mathrm{and} \ 
    \frac{1}{\sqrt{2}}\begin{bmatrix}
        0 & 0 & 1 \\
        0 & 0 & 0 \\
        1 & 0 & 0
    \end{bmatrix} \ \mathrm{and} \ 
    \frac{1}{\sqrt{2}}\begin{bmatrix}
        0 & 1 & 0 \\
        1 & 0 & 0 \\
        0 & 0 & 0
    \end{bmatrix} \nonumber
\end{equation}
yielding a similar dependence (up to a factor) to $E_g$:
\begin{equation}
    I_{T_{2g}} = \frac{|\boldsymbol{\varepsilon}_i|^2|\boldsymbol{\varepsilon}_o|^2}{30}\times\frac{1}{10}R_{T_{2g}}^2\left(1 + \frac{1}{6} \cos^2{\alpha}\right) 
\end{equation}
$T_{1g}$ is described by off anti-symmetric diagonal matrices of type:
\begin{equation}
    \frac{1}{\sqrt{2}}\begin{bmatrix}
        0 & 0 & 0 \\
        0 & 0 & 1 \\
        0 & -1 & 0
    \end{bmatrix} \ \mathrm{and} \ 
    \frac{1}{\sqrt{2}}\begin{bmatrix}
        0 & 0 & 1 \\
        0 & 0 & 0 \\
        -1 & 0 & 0
    \end{bmatrix} \ \mathrm{and} \ 
    \frac{1}{\sqrt{2}}\begin{bmatrix}
        0 & 1 & 0 \\
        -1 & 0 & 0 \\
        0 & 0 & 0
    \end{bmatrix} \nonumber
\end{equation}
with subsequently:
\begin{equation}
    I_{T_{1g}} = \frac{|\boldsymbol{\varepsilon}_i|^2|\boldsymbol{\varepsilon}_o|^2}{30}\times \frac{1}{6}R_{T_{1g}}^2\left(1 - \frac{1}{2} \cos^2{\alpha}\right) 
\end{equation}
Note that for $T_{1g}$ and $T_{2g}$ representations, contrary to $E_g$, rotations of the standard basis will not lead to the same angular dependence.

The total intensity is the sum of all contributing final states, each described within an irrep and weighted by their respective dimensions:
\begin{eqnarray}
    I &=& I_{A_{1g}} + 2I_{E_g} + 3I_{T_{1g}} + 3I_{T_{2g}} 
\end{eqnarray}
The exact relationship between the various state coefficients ($R_{A_{1g}}$, $R_{T_{2g}}$, $R_{E_{g}}$, $R_{T_{1g}}$) depends on the specific final state and their approximated calculation is explicated in appendix~\ref{sec:AppendixB}. 

\section{$O_h$ spin-free $2p3d$ E1E1 transition polarization dependence} \label{sec:AppendixB}

This section gives the angular dependence of various final configurations involved in the L$_3$M$_5$ RIXS, i.e. $3d^95d_{e_g}^1$ and $3d^95d_{t_{2g}}^1$, both composed from several final states. 

Based on the decomposed tensor structure, it is possible to figure out the relation among factors $R$ under different irreps. The transition follows the configuration change $\ket{g} \to \ket{2p^54f^{0/1}5d^1} \to \ket{3d^94f^{0/1}5d^1}$. The transition tensor structure is totally determined by the electron configuration of $3d^95d^1$. The configuration $4f^{0/1}$ doesn't affect the irreps, because for $4f^{1}$ case, an additional electron-hole pair locates at two $a_{2u}$ ($f$-like) orbitals, as compared to  $4f^{0}$. Given $a_{2u}\times a_{2u}=A_{1g}$, this additional pair doesn't change the irrep of $3d^95d^1$ configuration.

We define the integral:
\begin{equation}
    |\bra{2p_x}x\ket{5d_{x^2-y^2}} \bra{3d_{xy}}y\ket{2p_{x}}| = R_{2p3d}
\end{equation}
for a particular choice of the polarization directions. All transition tensors can be scaled to $R_{2p3d}$. Assuming the excited electron locates at $5d_{x^2-y^2}$ and considering all intermediate states (2$p_x$, 2$p_y$ and 2$p_z$), the corresponding transition amplitude (for the same choice of polarization) is:
\begin{equation}
    \mathbf{D}_{gn} = |\bra{2p_x}x\ket{5d_{x^2-y^2}}|
    \begin{bmatrix}
        1 & 0 & 0 \\
        0 & -1 & 0 \\
        0 & 0 & 0
    \end{bmatrix}
    \begin{bmatrix}
        \ket{2p_x} \\ \ket{2p_y} \\ \ket{2p_z}
    \end{bmatrix}
\end{equation}
and similarly, assuming the final state hole is $3d_{xy}$, the corresponding transition amplitude:
\begin{equation}
    \mathbf{D}_{nf} = |\bra{3d_{xy}}y\ket{2p_x}|
    \begin{bmatrix}
        0 & 1 & 0 \\
        1 & 0 & 0 \\
        0 & 0 & 0
    \end{bmatrix}
    \begin{bmatrix}
        \ket{2p_x} \\ \ket{2p_y} \\ \ket{2p_z}
    \end{bmatrix}
\end{equation}
The E1E1 transition to the final state $3d_{xy} 5d_{x^2-y^2}^1$ is (similar to $3d^9$ with hole in $xy$):
\begin{equation}
    \mathbf{M} = \mathbf{D}_{gn} \otimes \mathbf{D}_{nf} = R_{2p3d} \begin{bmatrix}
        0 & 1 & 0 \\
        -1 & 0 & 0 \\
        0 & 0 & 0
    \end{bmatrix}
\end{equation}
a general result regardless of the polarization directions. All tensors can be calculated in the same approach. 

The $3d$ orbitals in $O_h$ belong to either $e_g$ or $t_{2g}$ and therefore the final states with the structure $3d^95d^1$ decompose to:
\begin{eqnarray}
&3d^95d_{e_g}^1:&(e_g \oplus t_{2g}) \otimes e_g  = (e_g\otimes e_g) \oplus  (e_g\otimes t_{2g}) \nonumber \\
&3d^95d_{t_{2g}}^1:&(e_g \oplus t_{2g}) \otimes t_{2g}  = (t_{2g}\otimes t_{2g}) \oplus  (e_g\otimes t_{2g}) \nonumber\\
\label{eq:decompose3d95d1}
\end{eqnarray}
Based on this, we calculate all transition tensors $\mathbf{N}$ between the different subspaces in Eq.~\ref{eq:decompose3d95d1}.

\begin{equation}
    \begin{array}{c|cc}
        e_g \otimes e_g & d_{x^2-y^2} & d_{z^2} \\ \hline

        d_{x^2-y^2}
        & \begin{bmatrix}
            1 & 0 & 0 \\
            0 & 1 & 0 \\
            0 & 0 & 0
        \end{bmatrix} &
        \begin{bmatrix}
            -\frac{1}{\sqrt{3}} & 0 & 0 \\
            0 & \frac{1}{\sqrt{3}} & 0 \\
            0 & 0 & 0
        \end{bmatrix}\\

        d_{z^2}
        & \begin{bmatrix}
            -\frac{1}{\sqrt{3}} & 0 & 0 \\
            0 & \frac{1}{\sqrt{3}} & 0 \\
            0 & 0 & 0
        \end{bmatrix}
        & \begin{bmatrix}
            \frac{1}{3} & 0 & 0 \\
            0 & \frac{1}{3} & 0 \\
            0 & 0 & \frac{4}{3}
        \end{bmatrix} 
    \end{array}
\end{equation}

\begin{equation}
    \begin{array}{c|cc}
         e_g\otimes t_{2g} & d_{x^2-y^2} & d_{z^2} \\ \hline

        d_{xy} 
        & \begin{bmatrix}
            0 & 1 & 0 \\
            -1 & 0 & 0 \\
            0 & 0 & 0
        \end{bmatrix}
        & \begin{bmatrix}
            0 & -\frac{1}{\sqrt{3}} & 0 \\
            -\frac{1}{\sqrt{3}} & 0 & 0 \\
            0 & 0 & 0
        \end{bmatrix} \\

        d_{yz}
        & \begin{bmatrix}
            0 & 0 & 0 \\
            0 & 0 & -1 \\
            0 & 0 & 0
        \end{bmatrix}
        & \begin{bmatrix}
            0 & 0 & 0 \\
            0 & 0 & \frac{2}{\sqrt{3}} \\
            0 & -\frac{1}{\sqrt{3}} & 0
        \end{bmatrix} \\

        d_{xz}
        & \begin{bmatrix}
            0 & 0 & 1 \\
            0 & 0 & 0 \\
            0 & 0 & 0
        \end{bmatrix}
        & \begin{bmatrix}
            0 & 0 & \frac{2}{\sqrt{3}} \\
            0 & 0 & 0 \\
            -\frac{1}{\sqrt{3}} & 0 & 0
        \end{bmatrix}
    \end{array}
\end{equation}

\begin{equation}
    \begin{array}{c|ccc}
        t_{2g}\otimes t_{2g} & d_{xy} & d_{yz} & d_{xz} \\ \hline

        d_{xy} 
        & \begin{bmatrix}
            1 & 0 & 0 \\
            0 & 1 & 0 \\
            0 & 0 & 0
        \end{bmatrix}
        & \begin{bmatrix}
            0 & 0 & 1 \\
            0 & 0 & 0 \\
            0 & 0 & 0
        \end{bmatrix}
        & \begin{bmatrix}
            0 & 0 & 0 \\
            0 & 0 & 1 \\
            0 & 0 & 0
        \end{bmatrix} \\

        d_{yz}
        & \begin{bmatrix}
            0 & 0 & 0 \\
            0 & 0 & 0 \\
            1 & 0 & 0
        \end{bmatrix}
        & \begin{bmatrix}
            0 & 0 & 0 \\
            0 & 1 & 0 \\
            0 & 0 & 1
        \end{bmatrix}
        & \begin{bmatrix}
            0 & 0 & 0 \\
            1 & 0 & 0 \\
            0 & 0 & 0
        \end{bmatrix} \\

        d_{xz}
        & \begin{bmatrix}
            0 & 0 & 0 \\
            0 & 0 & 0 \\
            0 & 1 & 0
        \end{bmatrix}
        & \begin{bmatrix}
            0 & 1 & 0 \\
            0 & 0 & 0 \\
            0 & 0 & 0
        \end{bmatrix}
        & \begin{bmatrix}
            1 & 0 & 0 \\
            0 & 0 & 0 \\
            0 & 0 & 1
        \end{bmatrix} 
    \end{array}
\end{equation}

The matrices of the irreps product form a new space. To get the final state structure, we need to decompose these spaces into subspaces represented by irreps. The irrep subspace can be calculated by projection. The Hilbert-Schmidt inner production of $\mathbf{N}$ on a base $\mathbf{B}$ is defined as following:
\begin{equation}
    P_{\mathbf{B}} = \Tr(\mathbf{N}^*\mathbf{B})
\end{equation}
where the bases $\mathbf{B}$ have already been shown in appendix~\ref{sec:AppendixA}. 

In the following we calculate the scaling coefficients $R / R_{2p3d}$, for each subspace.

$e_g \otimes e_g = A_{1g} \oplus A_{2g} \oplus E_{g}$:
\begin{equation}
    \begin{array}{c|cc}
        A_{1g} \subset e_g \otimes e_g & d_{x^2-y^2} & d_{z^2} \\ 
        \hline
        d_{x^2-y^2} & \frac{2}{\sqrt{3}} & 0 \\
        d_{z^2} & 0 & \frac{2}{\sqrt{3}}
    \end{array}
\end{equation}
\begin{eqnarray}
    R_{A_{1g}}^{e_g \otimes e_g} / R_{2p3d} &=& \sqrt{\left(\frac{2}{\sqrt{3}}\right)^2+\left(\frac{2}{\sqrt{3}}\right)^2} \nonumber\\
        &=& 2\sqrt{\frac{2}{3}}
\end{eqnarray}
For components with dimension larger than 1, e.g. $E_g$, we get the same coefficient values for any of the basis matrices $\mathbf{B}$, as is shown below:
\begin{equation}
    \begin{array}{c|cc}
        E_{g} \subset e_g \otimes e_g & d_{x^2-y^2} & d_{z^2} \\ 
        \hline
        d_{x^2-y^2} & 0, -\sqrt{\frac{2}{3}} & -\sqrt{\frac{2}{3}}, 0 \\
        d_{z^2} & -\sqrt{\frac{2}{3}}, 0 & 0, \sqrt{\frac{2}{3}}
    \end{array}
\end{equation}
\begin{eqnarray}
    R_{E_g}^{e_g \otimes e_g}/ R_{2p3d} &=& \sqrt{\left(-\sqrt{\frac{2}{3}}\right)^2+\left(-\sqrt{\frac{2}{3}}\right)^2} \nonumber\\
        &=& \sqrt{\left(-\sqrt{\frac{2}{3}}\right)^2+\left(\sqrt{\frac{2}{3}}\right)^2} \nonumber\\
        &=& \frac{2}{\sqrt{3}}
\end{eqnarray}
Since $A_{2g}$ is not an allowed final state symmetry by E1E1 transition, we do not further treat it here.

$e_g \otimes t_{2g} = T_{1g} \oplus T_{2g}$:
\begin{equation}
    \begin{array}{c|cc}
        T_{1g} \subset e_g \otimes t_{2g} & d_{x^2-y^2} & d_{z^2} \\ \hline
        d_{xy} & 0,0,\sqrt{2} & 0,0,0 \\
        d_{yz} & 0,\frac{1}{\sqrt{2}},0 & 0,-\sqrt{\frac{3}{2}},0 \\
        d_{xz} & -\frac{1}{\sqrt{2}},0,0 & -\sqrt{\frac{3}{2}},0,0 \\
    \end{array}
\end{equation}
\begin{eqnarray}
    R_{T_{1g}}^{e_g \otimes t_{2g}} / R_{2p3d} &=& \sqrt{(\sqrt{2})^2} \nonumber\\
        &=& \sqrt{\left(\frac{1}{\sqrt{2}}\right)^2+\left(-\sqrt{\frac{3}{2}}\right)^2} \nonumber\\
        &=& \sqrt{\left(-\frac{1}{\sqrt{2}}\right)^2+\left(-\sqrt{\frac{3}{2}}\right)^2} \nonumber\\
        &=& \sqrt{2} 
\end{eqnarray}

\begin{equation}
    \begin{array}{c|cc}
        T_{2g} \subset e_g \otimes t_{2g} & d_{x^2-y^2} & d_{z^2} \\ \hline
        d_{xy} & 0,0,0 & 0,0,-\sqrt{\frac{2}{3}} \\
        d_{yz} & 0,-\frac{1}{\sqrt{2}},0 & 0,\frac{1}{\sqrt{6}},0 \\
        d_{xz} & \frac{1}{\sqrt{2}},0,0 & -\frac{1}{\sqrt{6}},0,0 \\
    \end{array}
\end{equation}
\begin{eqnarray}
    R_{T_{2g}}^{e_g \otimes t_{2g}} / R_{2p3d} &=& \sqrt{\left(-\sqrt{\frac{2}{3}}\right)^2} \nonumber\\
        &=& \sqrt{\left(-\frac{1}{\sqrt{2}}\right)^2+\left(\frac{1}{\sqrt{6}}\right)^2} \nonumber\\
        &=& \sqrt{\left(\frac{1}{\sqrt{2}}\right)^2+\left(-\frac{1}{\sqrt{6}}\right)^2} \nonumber\\
        &=& \sqrt{\frac{2}{3}}   
\end{eqnarray}

$t_{2g} \otimes t_{2g} = A_{1g} \oplus E_{g} \oplus T_{1g} \oplus T_{2g}$:
\begin{equation}
    \begin{array}{c|ccc}
        A_{1g} \subset t_{2g}\otimes t_{2g} & d_{xy} & d_{yz} & d_{xz} \\ \hline
        d_{xy} & \frac{2}{\sqrt{3}} & 0 & 0 \\
        d_{yz} & 0 & \frac{2}{\sqrt{3}} & 0 \\
        d_{xz} & 0 & 0 & \frac{2}{\sqrt{3}} \\
    \end{array}
\end{equation}
\begin{eqnarray}
    R_{A_{1g}}^{t_{2g} \otimes t_{2g}} / R_{2p3d} &=& \sqrt{\left(\frac{2}{\sqrt{3}}\right)^2+\left(\frac{2}{\sqrt{3}}\right)^2+\left(\frac{2}{\sqrt{3}}\right)^2} \nonumber\\
        &=& 2
\end{eqnarray}

\begin{equation}
    \begin{array}{c|ccc}
        E_{g} \subset t_{2g}\otimes t_{2g} & d_{xy} & d_{yz} & d_{xz} \\ \hline
        d_{xy} & 0,-\sqrt{\frac{2}{3}} & 0,0 & 0,0 \\
        d_{yz} & 0,0 & -\frac{1}{\sqrt{2}},\frac{1}{\sqrt{6}} & 0,0 \\
        d_{xz} & 0,0 & 0,0 & \frac{1}{\sqrt{2}},\frac{1}{\sqrt{6}} \\
    \end{array}
\end{equation}
\begin{eqnarray}
    R_{E_{g}}^{t_{2g} \otimes t_{2g}} / R_{2p3d} &=& \sqrt{\left(\frac{1}{\sqrt{2}}\right)^2+\left(\frac{1}{\sqrt{2}}\right)^2} \nonumber\\
        &=& \sqrt{\left(-\sqrt{\frac{2}{3}}\right)^2+\left(\frac{1}{\sqrt{6}}\right)^2+\left(\frac{1}{\sqrt{6}}\right)^2} \nonumber\\
        &=& 1
\end{eqnarray}

\begin{equation}
    \begin{array}{c|ccc}
        T_{1g} \subset t_{2g}\otimes t_{2g} & d_{xy} & d_{yz} & d_{xz} \\ \hline
        d_{xy} & 0,0,0 & 0,-\frac{1}{\sqrt{2}},0 & -\frac{1}{\sqrt{2}},0,0 \\
        d_{yz} & 0,\frac{1}{\sqrt{2}},0 & 0,0,0 & 0,0,\frac{1}{\sqrt{2}} \\
        d_{xz} & \frac{1}{\sqrt{2}},0,0 & 0,0,-\frac{1}{\sqrt{2}} & 0,0,0 \\
    \end{array}
\end{equation}
\begin{eqnarray}
    R_{T_{1g}}^{t_{2g} \otimes t_{2g}} / R_{2p3d} &=& \sqrt{\left(\frac{1}{\sqrt{2}}\right)^2+\left(-\frac{1}{\sqrt{2}}\right)^2} \nonumber\\
        &=& \sqrt{\left(\frac{1}{\sqrt{2}}\right)^2+\left(-\frac{1}{\sqrt{2}}\right)^2} \nonumber\\
        &=& \sqrt{\left(\frac{1}{\sqrt{2}}\right)^2+\left(-\frac{1}{\sqrt{2}}\right)^2} \nonumber\\
        &=& 1
\end{eqnarray}

\begin{equation}
    \begin{array}{c|ccc}
        T_{2g} \subset t_{2g}\otimes t_{2g} & d_{xy} & d_{yz} & d_{xz} \\ \hline
        d_{xy} & 0,0,0 & 0,\frac{1}{\sqrt{2}},0 & \frac{1}{\sqrt{2}},0,0 \\
        d_{yz} & 0,\frac{1}{\sqrt{2}},0 & 0,0,0 & 0,0,\frac{1}{\sqrt{2}} \\
        d_{xz} & \frac{1}{\sqrt{2}},0,0 & 0,0,\frac{1}{\sqrt{2}} & 0,0,0 \\
    \end{array}
\end{equation}
\begin{eqnarray}
    R_{T_{2g}}^{t_{2g} \otimes t_{2g}} / R_{2p3d} &=& \sqrt{\left(\frac{1}{\sqrt{2}}\right)^2+\left(\frac{1}{\sqrt{2}}\right)^2} \nonumber\\
        &=& \sqrt{\left(\frac{1}{\sqrt{2}}\right)^2+\left(\frac{1}{\sqrt{2}}\right)^2} \nonumber\\
        &=& \sqrt{\left(\frac{1}{\sqrt{2}}\right)^2+\left(\frac{1}{\sqrt{2}}\right)^2} \nonumber\\
        &=& 1
\end{eqnarray}

The total intensity is therefore obtained by summing up the individual contributions of all irreps participating in the decomposition, weighted by their respective dimensions. Furthermore the individual terms need to be scaled, depending on the subspace the irrep originates from. For instance, $3d^95d_{e_g}^1$ has $T_{2g} \oplus T_{1g}$ from $t_{2g} \otimes e_{g}$ and $A_{1g} \oplus E_{g}$ from $e_{g} \otimes e_{g}$, we therefore need to consider the angular dependencies in appendix~\ref{sec:AppendixA} with the specific coefficients $R_{T_{1g}}^{t_{2g} \otimes e_{g}}, R_{T_{2g}}^{t_{2g} \otimes e_{g}}, R_{A_{1g}}^{e_{g} \otimes e_{g}}, R_{E_{g}}^{e_{g} \otimes e_{g}}$. After substitution:
\begin{eqnarray}
    I_{3d^95d_{e_g}^1} &=& I_{A_{1g}} + 2I_{E_g} + 3I_{T_{1g}} + 3I_{T_{2g}} \nonumber\\
    &=& \frac{|\boldsymbol{\varepsilon}_i|^2|\boldsymbol{\varepsilon}_o|^2}{30}\times \frac{66+\cos^2{\alpha}}{45} R_{2p3d}^2
\end{eqnarray}
For $3d^95d_{t_{2g}}^1$ we need to consider it contains $T_{2g}$ and $T_{1g}$ originating from both subspaces, therefore scaled distinctly. Finally:
\begin{eqnarray}
    I_{3d^95d_{t_{2g}}^1} &=& \frac{|\boldsymbol{\varepsilon}_i|^2|\boldsymbol{\varepsilon}_o|^2}{30}\times \frac{66+\cos^2{\alpha}}{30} R_{2p3d}^2
\end{eqnarray}

The spin-free vtc E1E1 transition polarization dependence can be deduced in a similar way.

In conclusion, we need to justify why including spin-orbit coupling (SOC) does not significantly alter the polarization dependencies derived in appendices~\ref{sec:AppendixA} and~\ref{sec:AppendixB}. For the considered transitions, each shell contains at most one electron or one hole. When SOC is included, each irrep splits into several related final states, contributing to different edges. SOC acts as a perturbation, causing only minor quantitative changes, primarily affecting the overall magnitude of the dependence rather than the relative weights of the irreps. Therefore, the polarization dependence remains qualitatively similar to the spin-free (SF) case even when SOC is included.

\section{Comparison between theory and \textit{ab initio} calculations}

This section compares the L$_3$M$_5$ RIXS irreps coefficients ratio from the group theory (analytical) to the \textit{ab initio} numerical results in OpenMolcas. 

\begin{table}[!htbp]
    \caption{Intensity ratio of SF and SOC (L$_{\alpha1}$) for the $5d_{e_g}$ peaks.}
    \begin{ruledtabular}
    \begin{tabular}{cccccc}
        Ratio & Theory & \multicolumn{2}{c}{$\ket{\mathrm{3d^94f^15d_{e_g}^1}}$} & \multicolumn{2}{c}{$\ket{\mathrm{3d^94f^05d_{e_g}^1}}$} \\ 
         & & SF & SOC & SF & SOC \\
        \hline
        $R_{A_{1g}}^2/R_{T_{2g}}^2$ & 4 & 3.63 & 3.76 & 3.61 & 3.62 \\
        $R_{E_{g}}^2/R_{T_{2g}}^2$ & 2 & 1.99 & 1.99 & 2.00 & 2.00 \\
        $R_{T_{1g}}^2/R_{T_{2g}}^2$ & 3 & 3.02 & 3.01 & 3.04 & 3.00
    \end{tabular}
    \end{ruledtabular}
    \label{tab:ratioctceg}
\end{table}

\begin{table}[!htbp]
    \caption{Intensity ratio of SF  $5d_{t_{2g}}$ peaks.}
    \begin{ruledtabular}
    \begin{tabular}{cccc}
        Ratio & Theory & $\ket{\mathrm{3d^94f^15d_{t_{2g}}^1}}$ & $\ket{\mathrm{3d^94f^05d_{t_{2g}}^1}}$ \\ 
        \hline
        $R_{A_{1g}}^2/R_{E_{g}}^2$ & 4 & 4.00 & 4.04 \\
        $R_{T_{1g}}^2/R_{E_{g}}^2$ & 3 & 2.96 & 3.06 \\
        $R_{T_{2g}}^2/R_{E_{g}}^2$ & 1.67 & 1.65 & 1.62 
    \end{tabular}
    \end{ruledtabular}
    \label{tab:ratioctct2g}
\end{table}

The expected values for the $\ket{3d^9 4f^{0/1} 5d^1}$ configurations are reported in Tabs.~\ref{tab:ratioctceg} and~\ref{tab:ratioctct2g}. When a single irrep contains more than one final state, their squared contributions sum up. We conclude that the theoretical (analytical) results are consistent with the numerical ones for both configurations. Owing to the rotational properties of the $E_g$ basis, the SOC states of the $5d_{e_g}^1$ configuration exhibit an angular dependence that is only slightly modified with respect to the SF case presented in Appendix~\ref{sec:AppendixA}. This is no longer the case for the $5d_{t_{2g}}^1$ configuration, where the coupling among different components within the $t_{2g}$ irrep leads to a substantial modification of the angular dependence of the individual irreps. Nevertheless, the calculations show that the \emph{total} intensity, i.e., after summing over all relevant irreps, does not change significantly between the SF and SOC scenarios.

\section{Equivalence to the spherical tensor method}

In this work, the XAS intensity expressions are given in Eq.~\ref{eq:sigmaD} and Eq.~\ref{eq:sigmaQ2}. Equivalent formulae are explicitly presented in Ref.~\cite{Brouder_Site_2008}, so no further derivation is required here. We instead demonstrate that the spherical-tensor approach developed in Ref.~\cite{Burrow_Angular_2026} is fully equivalent to the Cartesian-tensor formulation used to perform the orientational average of E1E1 RIXS.

The spherical-tensor treatment starts from the standard SO(3) decomposition of the inner product between the transition operator 
$\mathbf{M}$ and its complex conjugate. Following Ref.~\cite{Burrow_Angular_2026}, this decomposition yields three independent scalar coefficients, denoted $S_0$, $S_1$ and $S_2$:
\begin{equation}
    S_b = \sum_{imjn} X_{im,jn}(b) \times M_{im} M_{jn}^*
\end{equation}
where
\begin{eqnarray}
    X_{im, jn}(b) &=& \sum_{\beta=-b}^b (-1)^{b-\beta}\sum_{\mu\nu\mu'\nu'} \braket{1\mu'1\nu|b\beta}\braket{1\mu1\nu'|b\bar{\beta}} \nonumber\\
    && \times A_{i\mu'}A_{m\nu}A_{n\mu}A_{j\nu'}
\end{eqnarray}

The $A_{ij}$ is the transfer matrix element (projector) between Cartesian and spherical tensor ($\mu, \nu, \mu', \nu' = -1, 0, +1$):
\begin{equation}
    \mathbf{A} = \begin{bmatrix}
        \frac{1}{\sqrt{2}} & 0 & -\frac{1}{\sqrt{2}} \\
        \frac{i}{\sqrt{2}} & 0 & \frac{i}{\sqrt{2}} \\
        0 & 1 & 0
    \end{bmatrix}
\end{equation}
This decomposition gives rise to the following equations:
\begin{eqnarray}
    S_0 &=& \frac{1}{3} |\Tr(\mathbf{M})|^2 \nonumber\\ 
    S_1 &=& \frac{1}{2}(-\Tr(\mathbf{M}^\dagger\mathbf{M}) + \Tr(\mathbf{M}^*\mathbf{M})) \nonumber\\
    S_2 &=& \frac{1}{2}(\Tr(\mathbf{M}^\dagger\mathbf{M}) + \Tr(\mathbf{M}^*\mathbf{M})) - \frac{1}{3}|\Tr(\mathbf{M})|^2 \label{eq:Sterms}
\end{eqnarray}

These three terms are separately the isotropic invariant, antisymmetric invariant and symmetric traceless invariant. The relationship between these invariants and intensity is given by Ref.~\cite{Juhin_Angular_2014}:
\begin{eqnarray}
    \braket{\sigma^{\mathrm{E1E1}}_{\mathrm{RIXS}}} &=& \sum_{g=0}^2 S_g \cdot \left[\frac{(-1)^g}{9}- \left\{\begin{matrix}
        1 & 1 & g \\ 1 & 1 & 1
    \end{matrix}\right\}
    \frac{|\boldsymbol{\hat{\varepsilon}}_i\cdot\boldsymbol{\hat{\varepsilon}}_o^*|^2-|\boldsymbol{\hat{\varepsilon}}_i\cdot\boldsymbol{\hat{\varepsilon}}_o|^2}{2} \right. \nonumber\\
    &+& \left.
    \frac{4}{(2-g)!(3+g)!}\left(\frac{|\boldsymbol{\hat{\varepsilon}}_i\cdot\boldsymbol{\hat{\varepsilon}}_o^*|^2+|\boldsymbol{\hat{\varepsilon}}_i\cdot\boldsymbol{\hat{\varepsilon}}_o|^2}{2}-\frac{1}{3}\right)\right]
    \nonumber
\end{eqnarray}

After reduction, the final intensity formula is:
\begin{eqnarray}
    \braket{\sigma^{\mathrm{E1E1}}_{\mathrm{RIXS}}} &=& -\frac{S_1}{6} + \frac{S_2}{10} + \left(\frac{S_0}{3}-\frac{S_2}{15}\right)|\boldsymbol{\hat{\varepsilon}}_i\cdot\boldsymbol{\hat{\varepsilon}}_o^*|^2 \nonumber\\ &&+ \left(\frac{S_1}{6} + \frac{S_2}{10}\right)|\boldsymbol{\hat{\varepsilon}}_i\cdot\boldsymbol{\hat{\varepsilon}}_o|^2 \label{eq:Isph}
\end{eqnarray}

After putting Eq.~\ref{eq:Sterms} into Eq.~\ref{eq:Isph}, the intensity formula become identical to the Eq.~\ref{eq:sigmaDD} obtained within our formalism.

%

\bibliographystyle{apsrev4-2}
\bibliography{ref}

@article{Nakazawa_Theory_2000,
author = {Nakazawa ,Makoto and Ogasawara ,Haruhiko and Kotani ,Akio},
title = {Theory of Polarization Dependence in Resonant X-Ray Emission Spectroscopy of Ce Compounds},
journal = {J. Phys. Soc. Jpn.},
volume = {69},
number = {12},
pages = {4071-4077},
year = {2000},
doi = {10.1143/JPSJ.69.4071},
URL = {https://doi.org/10.1143/JPSJ.69.4071},
}

@article{Joly_Resonant_2012,
  author    = {Joly, Y. and Di Matteo, S. and Bun{\u a}u, O.},
  title     = {Resonant X-ray diffraction: Basic theoretical principles},
  journal   = {Eur. Phys. J. Spec. Top.},
  volume    = {208},
  pages     = {21--38},
  year      = {2012},
  doi       = {10.1140/epjst/e2012-01604-5},
  url       = {https://link.springer.com/article/10.1140/epjst/e2012-01604-5}
}

@article{Brouder_Site_2008,
doi = {10.1088/0953-8984/20/45/455205},
url = {https://dx.doi.org/10.1088/0953-8984/20/45/455205},
year = {2008},
month = {oct},
volume = {20},
number = {45},
pages = {455205},
author = {Brouder, Christian and Juhin, Amélie and Bordage, Amélie and Arrio, Marie-Anne},
title = {Site symmetry and crystal symmetry: a spherical tensor analysis},
journal = {J. Phys.: Condens. Matter},
}

@article{Juhin_Angular_2014,
  author    = {Juhin, Amélie and Brouder, Christian and de Groot, Frank},
  title     = {Angular dependence of resonant inelastic x-ray scattering: a spherical tensor expansion},
  journal   = {Cent. Eur. J. Phys.},
  year      = {2014},
  volume    = {12},
  number    = {5},
  pages     = {323--340},
  doi       = {10.2478/s11534-014-0450-2},
  url       = {https://doi.org/10.2478/s11534-014-0450-2},
  issn      = {1644-3608},
}

@article{Ee_Combinatorics_2017,
doi = {10.1088/1361-6404/aa54ce},
url = {https://dx.doi.org/10.1088/1361-6404/aa54ce},
year = {2017},
month = {jan},
publisher = {IOP Publishing},
volume = {38},
number = {2},
pages = {025801},
author = {Ee, June-Haak and Jung, Dong-Won and Kim, U-Rae and Lee, Jungil},
title = {Combinatorics in tensor-integral reduction},
journal = {Eur. J. Phys.},
}

@article{Andrews_On_1977,
    author = {Andrews, D. L. and Thirunamachandran, T.},
    title = {On three‐dimensional rotational averages},
    journal = {J. Chem. Phys.},
    volume = {67},
    number = {11},
    pages = {5026-5033},
    year = {1977},
    month = {12},
    issn = {0021-9606},
    doi = {10.1063/1.434725},
    url = {https://doi.org/10.1063/1.434725},
}

@article{Andrews_Eighth_1981,
doi = {10.1088/0305-4470/14/6/008},
url = {https://dx.doi.org/10.1088/0305-4470/14/6/008},
year = {1981},
month = {jun},
publisher = {},
volume = {14},
number = {6},
pages = {1281},
author = {D L Andrews and W A Ghoul},
title = {Eighth rank isotropic tensors and rotational averages},
journal = {J. Phys. A: Math. Gen.}
}

@article{Groot_Resonant_2024,
  author  = {Frank M. F. de Groot and Maurits W. Haverkort and Hebatalla Elnaggar and Amélie Juhin and Ke-Jin Zhou and Pieter Glatzel},
  title   = {Resonant inelastic X-ray scattering},
  journal = {Nat. Rev. Methods Primers},
  year    = {2024},
  volume  = {4},
  number  = {1},
  pages   = {45},
  doi     = {10.1038/s43586-024-00322-6},
  url     = {https://doi.org/10.1038/s43586-024-00322-6},
  issn    = {2662-8449},
  month   = {jul}
}

@article{Samak_RIXS_2021,
title = {RIXS, XES and XAS studies for electronic structure of rare earth and alkaline earth modified manganite},
journal = {Physica B},
volume = {628},
pages = {413562},
year = {2022},
issn = {0921-4526},
doi = {https://doi.org/10.1016/j.physb.2021.413562},
url = {https://www.sciencedirect.com/science/article/pii/S092145262100716X},
author = {Mahmoud Abu-Samak and Upendra Kumar and A.M. Quraishi and Rajneesh Kumar and Shalendra Kumar and S. Dalela and M. Ayaz Ahmad and B.L. Choudhary and P.A. Alvi}
}

@article{Doring_Shake_2004,
  title = {Shake-up valence excitations in $\mathrm{CuO}$ by resonant inelastic x-ray scattering},
  author = {D\"oring, G. and Sternemann, C. and Kaprolat, A. and Mattila, A. and H\"am\"al\"ainen, K. and Sch\"ulke, W.},
  journal = {Phys. Rev. B},
  volume = {70},
  issue = {8},
  pages = {085115},
  numpages = {15},
  year = {2004},
  month = {Aug},
  publisher = {American Physical Society},
  doi = {10.1103/PhysRevB.70.085115},
  url = {https://link.aps.org/doi/10.1103/PhysRevB.70.085115}
}

@article{Couture_Polarization_2010,
  title = {Polarization dependence and symmetry analysis in indirect $K$-edge RIXS},
  author = {Chabot-Couture, G. and Hancock, J. N. and Mang, P. K. and Casa, D. M. and Gog, T. and Greven, M.},
  journal = {Phys. Rev. B},
  volume = {82},
  issue = {3},
  pages = {035113},
  numpages = {11},
  year = {2010},
  month = {Jul},
  publisher = {American Physical Society},
  doi = {10.1103/PhysRevB.82.035113},
  url = {https://link.aps.org/doi/10.1103/PhysRevB.82.035113}
}

@article{Kotani_Theory_2003,
    author = {Kotani, Akio},
    title = {Theory of Resonant Inelastic X‐ray Scattering in f and d Electron Systems},
    journal = {AIP Conf. Proc.},
    volume = {652},
    number = {1},
    pages = {338-346},
    year = {2003},
    month = {01},
    issn = {0094-243X},
    doi = {10.1063/1.1536394},
    url = {https://doi.org/10.1063/1.1536394},
}

@article{Rashid_Linear_2011,
  author    = {Muneer Ahmad Rashid and Faiz Ahmad and Naila Amir},
  title     = {Linear Invariants of a Cartesian Tensor Under SO(2), SO(3) and SO(4)},
  journal   = {Int. J. Theor. Phys.},
  year      = {2011},
  volume    = {50},
  number    = {2},
  pages     = {479--487},
  doi       = {10.1007/s10773-010-0555-3},
  url       = {https://doi.org/10.1007/s10773-010-0555-3},
  issn      = {1572-9575}
}

@article{Gordon_Orientation_2009,
doi = {10.1088/1742-6596/190/1/012047},
url = {https://dx.doi.org/10.1088/1742-6596/190/1/012047},
year = {2009},
month = {nov},
publisher = {},
volume = {190},
number = {1},
pages = {012047},
author = {R A Gordon and M W Haverkort and Subhra Sen Gupta and G A Sawatzky},
title = {Orientation-dependent x-ray Raman scattering from cubic crystals: Natural linear dichroism in MnO and CeO2},
journal = {J. Phys.: Conf. Ser.},
}

@article{Tegomo_Resonant_2022,
  title = {Resonant inelastic x-ray scattering of spin-charge excitations in a Kondo system},
  author = {Chiogo, B. Tegomo and Okamoto, J. and Li, J.-H. and Ohkochi, T. and Huang, H.-Y. and Huang, D.-J and Chen, C.-T. and Kuo, C.-N. and Lue, C.-S. and Chainani, A. and Malterre, D.},
  journal = {Phys. Rev. B},
  volume = {106},
  issue = {7},
  pages = {075141},
  numpages = {7},
  year = {2022},
  month = {Aug},
  publisher = {American Physical Society},
  doi = {10.1103/PhysRevB.106.075141},
  url = {https://link.aps.org/doi/10.1103/PhysRevB.106.075141}
}

@article{Kotani_Spectator_2012,
  author    = {Kotani, A. and Kvashnina, K. O. and Butorin, S. M. and Glatzel, P.},
  title     = {Spectator and participator processes in the resonant photon-in and photon-out spectra at the Ce L3 edge of CeO2},
  journal   = {Eur. Phys. J. B},
  year      = {2012},
  volume    = {85},
  number    = {8},
  pages     = {257},
  doi       = {10.1140/epjb/e2012-30079-1},
  url       = {https://doi.org/10.1140/epjb/e2012-30079-1},
  issn      = {1434-6036}
}

@article{Kang_Resolving_2019,
  title = {Resolving the nature of electronic excitations in resonant inelastic x-ray scattering},
  author = {Kang, M. and Pelliciari, J. and Krockenberger, Y. and Li, J. and McNally, D. E. and Paris, E. and Liang, R. and Hardy, W. N. and Bonn, D. A. and Yamamoto, H. and Schmitt, T. and Comin, R.},
  journal = {Phys. Rev. B},
  volume = {99},
  issue = {4},
  pages = {045105},
  numpages = {12},
  year = {2019},
  month = {Jan},
  publisher = {American Physical Society},
  doi = {10.1103/PhysRevB.99.045105},
  url = {https://link.aps.org/doi/10.1103/PhysRevB.99.045105}
}

@article{Veenendaal_Polarization_2006,
  title = {Polarization Dependence of $L$- and $M$-Edge Resonant Inelastic X-Ray Scattering in Transition-Metal Compounds},
  author = {van Veenendaal, Michel},
  journal = {Phys. Rev. Lett.},
  volume = {96},
  issue = {11},
  pages = {117404},
  numpages = {4},
  year = {2006},
  month = {Mar},
  publisher = {American Physical Society},
  doi = {10.1103/PhysRevLett.96.117404},
  url = {https://link.aps.org/doi/10.1103/PhysRevLett.96.117404}
}

@article{Krieger_Charge_2022,
  title = {Charge and Spin Order Dichotomy in ${\mathrm{NdNiO}}_{2}$ Driven by the Capping Layer},
  author = {Krieger, G. and Martinelli, L. and Zeng, S. and Chow, L. E. and Kummer, K. and Arpaia, R. and Moretti Sala, M. and Brookes, N. B. and Ariando, A. and Viart, N. and Salluzzo, M. and Ghiringhelli, G. and Preziosi, D.},
  journal = {Phys. Rev. Lett.},
  volume = {129},
  issue = {2},
  pages = {027002},
  numpages = {7},
  year = {2022},
  month = {Jul},
  publisher = {American Physical Society},
  doi = {10.1103/PhysRevLett.129.027002},
  url = {https://link.aps.org/doi/10.1103/PhysRevLett.129.027002}
}

@misc{Tagliavini_Polarization_2025,
      title={Polarization dependency in Resonant Inelastic X-Ray Scattering}, 
      author={Michelangelo Tagliavini and Fabian Wenzel and Maurits W. Haverkort},
      year={2025},
      eprint={2510.12891v1},
      archivePrefix={arXiv},
      primaryClass={cond-mat.mtrl-sci},
      url={https://arxiv.org/abs/2510.12891}, 
}

@article{Huang_Resonant_2022,
  author    = {Huang, H. Y. and Singh, A. and Wu, C. I. and Xie, J. D. and Okamoto, J. and Belik, A. A. and Kurmaev, E. and Fujimori, A. and Chen, C. T. and Streltsov, S. V. and Huang, D. J.},
  title     = {Resonant inelastic X-ray scattering as a probe of $J_{\mathrm{eff}} = 1/2$ state in 3d transition-metal oxide},
  journal   = {npj Quantum Mater.},
  year      = {2022},
  volume    = {7},
  number    = {1},
  pages     = {33},
  doi       = {10.1038/s41535-022-00430-0},
  url       = {https://doi.org/10.1038/s41535-022-00430-0}
}

@Article{Beheshti_In_2020,
author ="Beheshti Askari, Abbas and al Samarai, Mustafa and Hiraoka, Nozomu and Ishii, Hirofumi and Tillmann, Lukas and Muhler, Martin and DeBeer, Serena",
title  ="In situ X-ray emission and high-resolution X-ray absorption spectroscopy applied to Ni-based bimetallic dry methane reforming catalysts",
journal  ="Nanoscale",
year  ="2020",
volume  ="12",
issue  ="28",
pages  ="15185-15192",
publisher  ="The Royal Society of Chemistry",
doi  ="10.1039/D0NR01960G",
url  ="http://dx.doi.org/10.1039/D0NR01960G",
}

@article{Zasimov_HERFD_2022,
author = {Zasimov, Pavel and Amidani, Lucia and Retegan, Marius and Walter, Olaf and Caciuffo, Roberto and Kvashnina, Kristina O.},
title = {HERFD-XANES and RIXS Study on the Electronic Structure of Trivalent Lanthanides across a Series of Isostructural Compounds},
journal = {Inorg. Chem.},
volume = {61},
number = {4},
pages = {1817-1830},
year = {2022},
  doi     = {10.1021/acs.inorgchem.1c01525},
  url     = {https://doi.org/10.1021/acs.inorgchem.1c01525}
}

@article{Butorin_High_2016,
author = {Sergei M. Butorin  and Kristina O. Kvashnina  and Johan R. Vegelius  and Daniel Meyer  and David K. Shuh },
title = {High-resolution X-ray absorption spectroscopy as a probe of crystal-field and covalency effects in actinide compounds},
journal = {Proc. Natl. Acad. Sci. U.S.A.},
volume = {113},
number = {29},
pages = {8093-8097},
year = {2016},
doi = {10.1073/pnas.1601741113},
URL = {https://www.pnas.org/doi/abs/10.1073/pnas.1601741113}
}

@article{Burrow_Determination_2024,
author = {Burrow, Timothy G. and Alcock, Nathan M. and Huzan, Myron S. and Dunstan, Maja A. and Seed, John A. and Detlefs, Blanka and Glatzel, Pieter and Hunault, Myrtille O. J. Y. and Bendix, Jesper and Pedersen, Kasper S. and Baker, Michael L.},
title = {Determination of Uranium Central-Field Covalency with 3d4f Resonant Inelastic X-ray Scattering},
journal = {J. Am. Chem. Soc.},
volume = {146},
number = {32},
pages = {22570-22582},
year = {2024},
doi = {10.1021/jacs.4c06869},
URL = {https://doi.org/10.1021/jacs.4c06869}
}

@article{Sundermann_Resonant_2025,
  title = {Resonant inelastic x-ray scattering at the actinide ${\mathrm{M}}_{4,5}$-edges},
  author = {Sundermann, Martin and Hahn, Henrik and Christovam, Denise S. and Haverkort, Maurits W. and Caciuffo, Roberto and Keimer, Bernhard and Tjeng, Liu Hao and Severing, Andrea and Gretarsson, Hlynur},
  journal = {Phys. Rev. Res.},
  volume = {7},
  issue = {4},
  pages = {043081},
  numpages = {10},
  year = {2025},
  month = {Oct},
  publisher = {American Physical Society},
  doi = {10.1103/twsv-xh5j},
  url = {https://link.aps.org/doi/10.1103/twsv-xh5j}
}

@article{Glatzel_High_2005,
title = {High resolution 1s core hole X-ray spectroscopy in 3d transition metal complexes—electronic and structural information},
journal = {Coord. Chem. Rev.},
volume = {249},
number = {1},
pages = {65-95},
year = {2005},
issn = {0010-8545},
doi = {https://doi.org/10.1016/j.ccr.2004.04.011},
url = {https://www.sciencedirect.com/science/article/pii/S0010854504001146},
author = {Pieter Glatzel and Uwe Bergmann},
}

@article{Traulsen_The_2017,
doi = {10.1149/2.0091710jes},
url = {https://doi.org/10.1149/2.0091710jes},
year = {2017},
month = {jul},
publisher = {The Electrochemical Society},
volume = {164},
number = {10},
pages = {F3064},
author = {Traulsen, M. L. and de Carvalho, H. W. P. and Zielke, P. and Grunwaldt, J. -D.},
title = {The Effect of Electrical Polarization on Electronic Structure in LSM Electrodes: An Operando XAS, RIXS and XES Study},
journal = {J. Electrochem. Soc.},
}

@article{Amidani_Magnetic_2025,
  title = {Magnetic Exciton of EuS Revealed by Resonant Inelastic X-Ray Scattering},
  author = {Amidani, Lucia and Joos, Jonas J. and Glatzel, Pieter and Koloren\ifmmode \check{c}\else \v{c}\fi{}, Jind\ifmmode \check{r}\else \v{r}\fi{}ich},
  journal = {Phys. Rev. Lett.},
  volume = {134},
  issue = {4},
  pages = {046401},
  numpages = {7},
  year = {2025},
  month = {Jan},
  publisher = {American Physical Society},
  doi = {10.1103/PhysRevLett.134.046401},
  url = {https://link.aps.org/doi/10.1103/PhysRevLett.134.046401}
}

@article{Pedersen_Operando_2024,
Author = {Pedersen, Angus and Kumar, Kavita and Ku, Yu-Ping and Martin, Vincent
   and Dubau, Laetitia and Santos, Keyla Teixeira and Barrio, Jesus and
   Saveleva, Viktoriia A. and Glatzel, Pieter and Paidi, Vinod K. and Li,
   Xiaoyan and Hutzler, Andreas and Titirici, Maria-Magdalena and
   Bonnefont, Antoine and Cherevko, Serhiy and Stephens, Ifan E. L. and
   Maillard, Frederic},
Title = {Operando Fe dissolution in Fe-N-C electrocatalysts during acidic
   oxygen reduction: impact of local pH change},
Journal = {Energy Environ. Sci.},
Year = {2024},
Volume = {17},
Number = {17},
Pages = {6323-6337},
Month = {AUG 27},
DOI = {10.1039/d4ee01995d},
URL = {https://pubs.rsc.org/en/content/articlelanding/2024/ee/d4ee01995d}
}

@article{Orduz_L3_2024,
Author = {Orduz, Hugo Alexander Suarez and Bugarin, Luca and Heck, Sarina-Lena and
   Dolcet, Paolo and Casapu, Maria and Grunwaldt, Jan-Dierk and Glatzel,
   Pieter},
Title = {L3-edge X-ray spectroscopy of rhodium and palladium
   compounds},
Journal = {J. Synchrotron Radiat.},
Year = {2024},
Volume = {31},
Number = {4},
Pages = {733-740},
Month = {JUL},
DOI = {10.1107/S1600577524004673},
ISSN = {0909-0495},
EISSN = {1600-5775},
ResearcherID-Numbers = {Casapu, Maria/P-6861-2018
   Grunwaldt, Jan-Dierk/L-9246-2013
   Dolcet, Paolo/JDD-7353-2023
   Glatzel, Pieter/E-9958-2010},
Unique-ID = {WOS:001274892400011},
}

@article{Longo_Dynamic_2022,
Author = {Longo, Alessandro and Giannici, Francesco and Casaletto, Maria Pia and
   Rovezzi, Mauro and Sahle, Christoph J. and Glatzel, Pieter and Joly,
   Yves and Martorana, Antonino},
Title = {Dynamic Role of Gold d-Orbitals during CO Oxidation under Aerobic
   Conditions},
Journal = {ACS Catal.},
Year = {2022},
Volume = {12},
Number = {6},
Pages = {3615-3627},
Month = {MAR 18},
DOI = {10.1021/acscatal.1c05739},
URL = {https://pubs.acs.org/doi/10.1021/acscatal.1c05739}
}

@article{ Svyazhin_Chemical_2022,
Author = {Svyazhin, Artem and Nalbandyan, Vladimir and Rovezzi, Mauro and
   Chumakova, Aleksandra and Detlefs, Blanka and Guda, Alexander A. and
   Santambrogio, Alessandro and Manceau, Alain and Glatzel, Pieter},
Title = {Chemical Information in the L3 X-ray Absorption Spectra of
   Molybdenum Compounds by High-Energy-Resolution Detection and Density
   Functional Theory},
Journal = {Inorg. Chem.},
Year = {2022},
Volume = {61},
Number = {2},
Pages = {869-881},
Month = {JAN 17},
DOI = {10.1021/acs.inorgchem.1c02600},
URL = {https://pubs.acs.org/doi/10.1021/acs.inorgchem.1c02600}
}

@Article{Guo_HERFD_2020,
author ="Guo, Meiyuan and Prakash, Om and Fan, Hao and de Groot, Lisa H. M. and Hlynsson, Valtýr Freyr and Kaufhold, Simon and Gordivska, Olga and Velásquez, Nicolás and Chabera, Pavel and Glatzel, Pieter and Wärnmark, Kenneth and Persson, Petter and Uhlig, Jens",
title  ="HERFD-XANES probes of electronic structures of ironII/III carbene complexes",
journal  ="Phys. Chem. Chem. Phys.",
year  ="2020",
volume  ="22",
issue  ="16",
pages  ="9067-9073",
publisher  ="The Royal Society of Chemistry",
doi  ="10.1039/C9CP06309A",
url  ="http://dx.doi.org/10.1039/C9CP06309A",
}

@article{Vedrine_Heterogeneous_2017,
author = {Védrine, Jacques C.},
title = {Heterogeneous Catalysis on Metal Oxides},
journal = {Catalysts},
volume = {7},
YEAR = {2017},
number = {11},
pages = {341},
DOI = {10.3390/catal7110341},
URL = {https://www.mdpi.com/2073-4344/7/11/341},
ISSN = {2073-4344}
}

@article{deGroot_Multiplet_2005,
  author  = {de Groot, Frank M. F.},
  title   = {Multiplet effects in X-ray spectroscopy},
  journal = {Coord. Chem. Rev.},
  year    = {2005},
  volume  = {249},
  number  = {1--2},
  pages   = {31--63},
  doi     = {10.1016/j.ccr.2004.03.018},
  url     = {https://doi.org/10.1016/j.ccr.2004.03.018}
}

@article{Kotani_Resonant_2001,
  author  = {Kotani, Akio and Shin, Shik},
  title   = {Resonant inelastic X-ray scattering spectra for electrons in solids},
  journal = {Rev. Mod. Phys.},
  year    = {2001},
  volume  = {73},
  number  = {1},
  pages   = {203--246},
  doi     = {10.1103/RevModPhys.73.203},
  url     = {https://doi.org/10.1103/RevModPhys.73.203}
}

@article{Li_The_2023,
author = {Li Manni, Giovanni and Fdez. Galván, Ignacio and Alavi, Ali and Aleotti, Flavia and Aquilante, Francesco and Autschbach, Jochen and Avagliano, Davide and Baiardi, Alberto and Bao, Jie J. and Battaglia, Stefano and Birnoschi, Letitia and Blanco-González, Alejandro and Bokarev, Sergey I. and Broer, Ria and Cacciari, Roberto and Calio, Paul B. and Carlson, Rebecca K. and Carvalho Couto, Rafael and Cerdán, Luis and Chibotaru, Liviu F. and Chilton, Nicholas F. and Church, Jonathan Richard and Conti, Irene and Coriani, Sonia and Cu{\'e}llar-Zuquin, Juliana and Daoud, Razan E. and Dattani, Nike and Decleva, Piero and de Graaf, Coen and Delcey, Micka{\"e}l G. and De Vico, Luca and Dobrautz, Werner and Dong, Sijia S. and Feng, Rulin and Ferr{\'e}, Nicolas and Filatov(Gulak), Michael and Gagliardi, Laura and Garavelli, Marco and González, Leticia and Guan, Yafu and Guo, Meiyuan and Hennefarth, Matthew R. and Hermes, Matthew R. and Hoyer, Chad E. and Huix-Rotllant, Miquel and Jaiswal, Vishal Kumar and Kaiser, Andy and Kaliakin, Danil S. and Khamesian, Marjan and King, Daniel S. and Kochetov, Vladislav and Krośnicki, Marek and Kumaar, Arpit Arun and Larsson, Ernst D. and Lehtola, Susi and Lepetit, Marie-Bernadette and Lischka, Hans and López Ríos, Pablo and Lundberg, Marcus and Ma, Dongxia and Mai, Sebastian and Marquetand, Philipp and Merritt, Isabella C. D. and Montorsi, Francesco and M{\"o}rchen, Maximilian and Nenov, Artur and Nguyen, Vu Ha Anh and Nishimoto, Yoshio and Oakley, Meagan S. and Olivucci, Massimo and Oppel, Markus and Padula, Daniele and Pandharkar, Riddhish and Phung, Quan Manh and Plasser, Felix and Raggi, Gerardo and Rebolini, Elisa and Reiher, Markus and Rivalta, Ivan and Roca-Sanjuán, Daniel and Romig, Thies and Safari, Arta Anushirwan and Sánchez-Mansilla, Aitor and Sand, Andrew M. and Schapiro, Igor and Scott, Thais R. and Segarra-Martí, Javier and Segatta, Francesco and Sergentu, Dumitru-Claudiu and Sharma, Prachi and Shepard, Ron and Shu, Yinan and Staab, Jakob K. and Straatsma, Tjerk P. and Sørensen, Lasse Kragh and Tenorio, Bruno Nunes Cabral and Truhlar, Donald G. and Ungur, Liviu and Vacher, Morgane and Veryazov, Valera and Voß, Torben Arne and Weser, Oskar and Wu, Dihua and Yang, Xuchun and Yarkony, David and Zhou, Chen and Zobel, J. Patrick and Lindh, Roland},
title = {The OpenMolcas Web: A Community-Driven Approach to Advancing Computational Chemistry},
journal = {J. Chem. Theory Comput.},
volume = {19},
number = {20},
pages = {6933-6991},
year = {2023},
doi = {10.1021/acs.jctc.3c00182},
URL = {https://doi.org/10.1021/acs.jctc.3c00182}
}

@article{Polly_Relativistic_2021,
author = {Polly, Robert and Schacherl, Bianca and Rothe, J{\"o}rg and Vitova, Tonya},
title = {Relativistic Multiconfigurational Ab Initio Calculation of Uranyl 3d4f Resonant Inelastic X-ray Scattering},
journal = {Inorg. Chem.},
volume = {60},
number = {24},
pages = {18764-18776},
year = {2021},
doi = {10.1021/acs.inorgchem.1c02364},
URL = {https://doi.org/10.1021/acs.inorgchem.1c02364},
}

@book{Sakurai_Advanced_1967,
  author    = {Sakurai, Jun John},
  title     = {Advanced Quantum Mechanics},
  publisher = {Addison-Wesley},
  address   = {Reading, Massachusetts},
  year      = {1967},
  isbn      = {9780201067101}
}

@article{Ament_Resonant_2011,
  title = {Resonant inelastic x-ray scattering studies of elementary excitations},
  author = {Ament, Luuk J. P. and van Veenendaal, Michel and Devereaux, Thomas P. and Hill, John P. and van den Brink, Jeroen},
  journal = {Rev. Mod. Phys.},
  volume = {83},
  issue = {2},
  pages = {705--767},
  numpages = {0},
  year = {2011},
  month = {Jun},
  publisher = {American Physical Society},
  doi = {10.1103/RevModPhys.83.705},
  url = {https://link.aps.org/doi/10.1103/RevModPhys.83.705}
}

@article{Neese_The_2012,
author = {Neese, F.},
title = {The ORCA program system},
journal = {WIRES Comput. Molec. Sci.},
volume = {2},
number = {1},
pages = {73-78},
DOI = {10.1002/wcms.81},
year = {2012},
type = {journal Article}
}

@article{Fdez_OpenMolcas_2019,
author = {Fdez. Galv{\'a}n, Ignacio and Vacher, Morgane and Alavi, Ali and Angeli, Celestino and Aquilante, Francesco and Autschbach, Jochen and Bao, Jie J. and Bokarev, Sergey I. and Bogdanov, Nikolay A. and Carlson, Rebecca K. and Chibotaru, Liviu F. and Creutzberg, Joel and Dattani, Nike and Delcey, Micka{\"e}l G. and Dong, Sijia S. and Dreuw, Andreas and Freitag, Leon and Frutos, Luis Manuel and Gagliardi, Laura and Gendron, Fr{\'e}d{\'e}ric and Giussani, Angelo and Gonz{\'a}lez, Leticia and Grell, Gilbert and Guo, Meiyuan and Hoyer, Chad E. and Johansson, Marcus and Keller, Sebastian and Knecht, Stefan and Kova{\v c}evi{\'c}, Goran and K{\"a}llman, Erik and Li Manni, Giovanni and Lundberg, Marcus and Ma, Yingjin and Mai, Sebastian and Malhado, Jo{\~a}o Pedro and Malmqvist, Per {\AA}ke and Marquetand, Philipp and Mewes, Stefanie A. and Norell, Jesper and Olivucci, Massimo and Oppel, Markus and Phung, Quan Manh and Pierloot, Kristine and Plasser, Felix and Reiher, Markus and Sand, Andrew M. and Schapiro, Igor and Sharma, Prachi and Stein, Christopher J. and S{\o}rensen, Lasse Kragh and Truhlar, Donald G. and Ugandi, Mihkel and Ungur, Liviu and Valentini, Alessio and Vancoillie, Steven and Veryazov, Valera and Weser, Oskar and Weso{\l}owski, Tomasz A. and Widmark, Per-Olof and Wouters, Sebastian and Zech, Alexander and Zobel, J. Patrick and Lindh, Roland},
title = {{OpenMolcas}: From Source Code to Insight},
journal = {J. Chem. Theory Comput.},
volume = {15},
number = {11},
pages = {5925-5964},
year = {2019},
doi = {10.1021/acs.jctc.9b00532},
URL = {https://doi.org/10.1021/acs.jctc.9b00532}
}

@article{Aquilante_Modern_2020,
    author = {Aquilante, Francesco and Autschbach, Jochen and Baiardi, Alberto and Battaglia, Stefano and Borin, Veniamin A. and Chibotaru, Liviu F. and Conti, Irene and De Vico, Luca and Delcey, Micka\"{e}l and Fdez. Galv\'{a}n, Ignacio and Ferr\'{e}, Nicolas and Freitag, Leon and Garavelli, Marco and Gong, Xuejun and Knecht, Stefan and Larsson, Ernst D. and Lindh, Roland and Lundberg, Marcus and Malmqvist, Per \AA{}ke and Nenov, Artur and Norell, Jesper and Odelius, Michael and Olivucci, Massimo and Pedersen, Thomas B. and Pedraza-Gonz\'{a}lez, Laura and Phung, Quan M. and Pierloot, Kristine and Reiher, Markus and Schapiro, Igor and Segarra-Mart\'{i}, Javier and Segatta, Francesco and Seijo, Luis and Sen, Saumik and Sergentu, Dumitru-Claudiu and Stein, Christopher J. and Ungur, Liviu and Vacher, Morgane and Valentini, Alessio and Veryazov, Valera},
    title = {Modern quantum chemistry with {[Open]Molcas}},
    journal = {J. Chem. Phys.},
    volume = {152},
    number = {21},
    pages = {214117},
    year = {2020},
    month = {06},
    issn = {0021-9606},
    doi = {10.1063/5.0004835},
    url = {https://doi.org/10.1063/5.0004835}
}

@article{Derenzo_Determining_2000,
title = {Determining point charge arrays that produce accurate ionic crystal fields for atomic cluster calculations},
volume = {112},
issn = {0021-9606, 1089-7690},
url = {https://pubs.aip.org/jcp/article/112/5/2074/349106/Determining-point-charge-arrays-that-produce},
doi = {10.1063/1.480776},
language = {en},
number = {5},
journal = {J. Chem. Phys.},
author = {Derenzo, S. and Klintenberg, M. and Weber, M.},
year = {2000},
pages = {2074--2081},
}

@article{Klintenberg_Accurate_2000,
title = {Accurate crystal fields for embedded cluster calculations},
volume = {131},
copyright = {https://www.elsevier.com/tdm/userlicense/1.0/},
issn = {0010-4655},
url = {https://linkinghub.elsevier.com/retrieve/pii/S0010465500000710},
doi = {10.1016/s0010-4655(00)00071-0},
language = {en},
number = {1-2},
urldate = {2025-05-01},
journal = {Comput. Phys. Commun.},
author = {Klintenberg, M. and Derenzo, S. and Weber, M.},
year = {2000},
pages = {120--128},
}

@article{Pascual_Ab_1993,
    author = {Pascual, José Luis and Seijo, Luis and Barandiarán, Zoila},
    title = {Ab initio model potential study of environmental effects on the Jahn-Teller parameters of $\mathrm{Cu}^{2+}$ and $\mathrm{Ag}^{2+}$ impurities in {MgO}, {CaO}, and {SrO} hosts},
    journal = {J. Chem. Phys.},
    volume = {98},
    number = {12},
    pages = {9715-9724},
    year = {1993},
    month = {06},
    issn = {0021-9606},
    doi = {10.1063/1.464350},
    url = {https://doi.org/10.1063/1.464350},
}

@article{Nygren_Bonding_1994,
    author = {Nygren, Martin A. and Pettersson, Lars G. M. and Barandiarán, Zoila and Seijo, Luis},
    title = {Bonding between {CO} and the {MgO(001)} surface: A modified picture},
    journal = {J. Chem. Phys.},
    volume = {100},
    number = {3},
    pages = {2010-2018},
    year = {1994},
    month = {02},
    issn = {0021-9606},
    doi = {10.1063/1.466553},
    url = {https://doi.org/10.1063/1.466553}
}

@article{Gendron_Puzzling_2017,
author = {Gendron, Fr{\'e}d{\'e}ric and Autschbach, Jochen},
title = {Puzzling Lack of Temperature Dependence of the $\mathrm{PuO}_2$ Magnetic Susceptibility Explained According to Ab Initio Wave Function Calculations},
journal = {J. Phys. Chem. Lett.},
volume = {8},
number = {3},
pages = {673-678},
year = {2017},
doi = {10.1021/acs.jpclett.6b02968},
URL = {https://doi.org/10.1021/acs.jpclett.6b02968}
}

@inbook{Roos_The_1987,
author = {Roos, Björn O.},
publisher = {John Wiley \& Sons, Ltd},
isbn = {9780470142943},
title = {The Complete Active Space Self-Consistent Field Method and its Applications in Electronic Structure Calculations},
booktitle = {Advances in Chemical Physics},
chapter = {69},
pages = {399-445},
doi = {https://doi.org/10.1002/9780470142943.ch7},
url = {https://onlinelibrary.wiley.com/doi/abs/10.1002/9780470142943.ch7},
year = {1987},
}

@article{Roos_A_1980,
title = {A complete active space SCF method (CASSCF) using a density matrix formulated super-CI approach},
journal = {Chem. Phys.},
volume = {48},
number = {2},
pages = {157-173},
year = {1980},
issn = {0301-0104},
doi = {https://doi.org/10.1016/0301-0104(80)80045-0},
url = {https://www.sciencedirect.com/science/article/pii/0301010480800450},
author = {Björn O. Roos and Peter R. Taylor and Per E.M. Sigbahn},
}

@article{Siegbahn_A_1980,
doi = {10.1088/0031-8949/21/3-4/014},
url = {https://dx.doi.org/10.1088/0031-8949/21/3-4/014},
year = {1980},
month = {jan},
volume = {21},
number = {3-4},
pages = {323},
author = {Per Siegbahn and Anders Heiberg and Björn Roos and Bernard Levy},
title = {A Comparison of the Super-CI and the Newton-Raphson Scheme in the Complete Active Space SCF Method},
journal = {Phys. Scr.},
}

@article{Siegbahn_The_1981,
    author = {Siegbahn, Per E. M. and Almlöf, Jan and Heiberg, Anders and Roos, Björn O.},
    title = {The complete active space SCF (CASSCF) method in a Newton–Raphson formulation with application to the HNO molecule},
    journal = {J. Chem. Phys.},
    volume = {74},
    number = {4},
    pages = {2384-2396},
    year = {1981},
    month = {02},
    issn = {0021-9606},
    doi = {10.1063/1.441359},
    url = {https://doi.org/10.1063/1.441359}
}

@article{Malmqvist_The_1990,
author = {Malmqvist, Per Aake. and Rendell, Alistair. and Roos, Bjoern O.},
title = {The restricted active space self-consistent-field method, implemented with a split graph unitary group approach},
journal = {J. Phys. Chem.},
volume = {94},
number = {14},
pages = {5477-5482},
year = {1990},
doi = {10.1021/j100377a011},
URL = {https://doi.org/10.1021/j100377a011}
}

@article{Douglas_Quantum_1974,
title = {Quantum electrodynamical corrections to the fine structure of helium},
journal = {Ann. Phys.},
volume = {82},
number = {1},
pages = {89-155},
year = {1974},
issn = {0003-4916},
doi = {https://doi.org/10.1016/0003-4916(74)90333-9},
url = {https://www.sciencedirect.com/science/article/pii/0003491674903339},
author = {Marvin Douglas and Norman M Kroll},
}

@article{Hess_Applicability_1985,
  title = {Applicability of the no-pair equation with free-particle projection operators to atomic and molecular structure calculations},
  author = {Hess, Bernd A.},
  journal = {Phys. Rev. A},
  volume = {32},
  issue = {2},
  pages = {756--763},
  year = {1985},
  month = {Aug},
  publisher = {American Physical Society},
  doi = {10.1103/PhysRevA.32.756},
  url = {https://link.aps.org/doi/10.1103/PhysRevA.32.756}
}

@article{Hess_Relativistic_1986,
  title = {Relativistic electronic-structure calculations employing a two-component no-pair formalism with external-field projection operators},
  author = {Hess, Bernd A.},
  journal = {Phys. Rev. A},
  volume = {33},
  issue = {6},
  pages = {3742--3748},
  year = {1986},
  month = {Jun},
  publisher = {American Physical Society},
  doi = {10.1103/PhysRevA.33.3742},
  url = {https://link.aps.org/doi/10.1103/PhysRevA.33.3742}
}

@article{Wolf_The_2002,
    author = {Wolf, Alexander and Reiher, Markus and Hess, Bernd Artur},
    title = {The generalized Douglas–Kroll transformation},
    journal = {J. Chem. Phys.},
    volume = {117},
    number = {20},
    pages = {9215-9226},
    year = {2002},
    month = {11},
    issn = {0021-9606},
    doi = {10.1063/1.1515314},
    url = {https://doi.org/10.1063/1.1515314}
}

@article{Pedersen_Density_2009,
  author       = {Thomas Bondo Pedersen and Francesco Aquilante and Roland Lindh},
  title        = {Density fitting with auxiliary basis sets from Cholesky decompositions},
  journal      = {Theor. Chem. Acc.},
  year         = {2009},
  volume       = {124},
  number       = {1},
  pages        = {1--10},
  doi          = {10.1007/s00214-009-0608-y},
  url          = {https://doi.org/10.1007/s00214-009-0608-y},
  issn         = {1432-2234},
}

@article{Dyall_Finite_1993,
title = {Finite nucleus effects on relativistic energy corrections},
journal = {Chem. Phys. Lett.},
volume = {201},
number = {1},
pages = {27-32},
year = {1993},
issn = {0009-2614},
doi = {https://doi.org/10.1016/0009-2614(93)85028-M},
url = {https://www.sciencedirect.com/science/article/pii/000926149385028M},
author = {Kenneth G. Dyall and Knut Fægri},
}

@article{Malmqvist_The_2002,
title = {The restricted active space {(RAS)} state interaction approach with spin–orbit coupling},
journal = {Chem. Phys. Lett.},
volume = {357},
number = {3},
pages = {230-240},
year = {2002},
issn = {0009-2614},
doi = {https://doi.org/10.1016/S0009-2614(02)00498-0},
url = {https://www.sciencedirect.com/science/article/pii/S0009261402004980},
author = {Per Åke Malmqvist and Björn O. Roos and Bernd Schimmelpfennig},
}

@article{Kummerle_The_1999,
title = {The Structures of {C-$\mathrm{Ce_{2}O_{3+\delta}}$}, $\mathrm{Ce_{7}O_{12}}$, and $\mathrm{Ce_{11}O_{20}}$},
journal = {J. Solid State Chem.},
volume = {147},
number = {2},
pages = {485-500},
year = {1999},
issn = {0022-4596},
doi = {https://doi.org/10.1006/jssc.1999.8403},
url = {https://www.sciencedirect.com/science/article/pii/S0022459699984037},
author = {E.A Kümmerle and G Heger},
}

@misc{Polarixs,
  author = {Sihan Zhang},
  title = {{Polarixs}: A Python package for angular and polarization dependent RIXS convolution},
  howpublished = {\url{https://github.com/ZhangHMDS/Polarixs}},
  note = {Accessed: 2025-08-27},
  year = {2025}
}

@misc{DAXS,
  author = {Marius Retegan},
  title = {{DAXS}: Data Analysis for X-ray Spectroscopy},
  howpublished = {\url{https://gitlab.esrf.fr/spectroscopy/daxs}},
  note = {Accessed: 2025-10-20},
  year = {2025}
}

@article{Glatzel_The_2021,
author = "Glatzel, Pieter and Harris, Alistair and Marion, Philippe and Sikora, Marcin and Weng, Tsu-Chien and Guilloud, Cyril and Lafuerza, Sara and Rovezzi, Mauro and Detlefs, Blanka and Ducott{\'{e}}, Ludovic",
title = "{The five-analyzer point-to-point scanning crystal spectrometer at ESRF ID26}",
journal = "J. Synchrotron Rad.",
year = "2021",
volume = "28",
number = "1",
pages = "362--371",
month = "Jan",
doi = {10.1107/S1600577520015416},
url = {https://doi.org/10.1107/S1600577520015416}
}

@article{Sergentu_Probing_2021,
author = {Sergentu, Dumitru-Claudiu and Booth, Corwin H. and Autschbach, Jochen},
title = {Probing Multiconfigurational States by Spectroscopy: The Cerium XAS L3-edge Puzzle},
journal = {Chem. Eur. J},
volume = {27},
number = {25},
pages = {7239-7251},
doi = {https://doi.org/10.1002/chem.202100145},
url = {https://chemistry-europe.onlinelibrary.wiley.com/doi/abs/10.1002/chem.202100145},
year = {2021}
}

@Article{Sergentu_Xray_2022,
author ="Sergentu, Dumitru-Claudiu and Autschbach, Jochen",
title  ="X-ray absorption spectra of f-element complexes: insight from relativistic multiconfigurational wavefunction theory",
journal  ="Dalton Trans.",
year  ="2022",
volume  ="51",
issue  ="5",
pages  ="1754-1764",
publisher  ="The Royal Society of Chemistry",
doi  ="10.1039/D1DT04075H",
url  ="http://dx.doi.org/10.1039/D1DT04075H",
}

@article{Brink_Theory_2005,
title = {Theory of indirect resonant inelastic X-ray scattering},
journal = {J. Phys. Chem. Solids},
volume = {66},
number = {12},
pages = {2145-2149},
year = {2005},
issn = {0022-3697},
doi = {https://doi.org/10.1016/j.jpcs.2005.10.168},
url = {https://www.sciencedirect.com/science/article/pii/S0022369705005706},
author = {Jeroen {van den Brink} and Michel {van Veenendaal}},
}

@article{Bunau_Self_2009,
doi = {10.1088/0953-8984/21/34/345501},
url = {https://doi.org/10.1088/0953-8984/21/34/345501},
year = {2009},
month = {aug},
publisher = {},
volume = {21},
number = {34},
pages = {345501},
author = {Bunău, O and Joly, Y},
title = {Self-consistent aspects of x-ray absorption calculations},
journal = {J. Phys.: Condens. Matter},
}

@article{Peacock_Natural_2001,
author = {Peacock, Robert D. and Stewart, Brian},
title = {Natural Circular Dichroism in X-ray Spectroscopy},
journal = {J. Phys. Chem. B},
volume = {105},
number = {2},
pages = {351-360},
year = {2001},
doi = {10.1021/jp001946y},
URL = {https://doi.org/10.1021/jp001946y},
}

@article{Carra_Xray_2003,
  title = {X-ray dichroism in noncentrosymmetric crystals},
  author = {Carra, Paolo and Jerez, Andr\'es and Marri, Ivan},
  journal = {Phys. Rev. B},
  volume = {67},
  issue = {4},
  pages = {045111},
  numpages = {8},
  year = {2003},
  month = {Jan},
  publisher = {American Physical Society},
  doi = {10.1103/PhysRevB.67.045111},
  url = {https://link.aps.org/doi/10.1103/PhysRevB.67.045111}
}

@article{Burrow_Angular_2026,
  title = {Angular dependence and powder average of resonant inelastic x-ray scattering},
  author = {Burrow, Timothy G. and Hunault, Myrtille O. J. Y. and Besnard, Fabien and Juhin, Am\'elie and Brouder, Christian},
  journal = {Phys. Rev. B},
  volume = {113},
  issue = {11},
  pages = {115131},
  numpages = {23},
  year = {2026},
  month = {Mar},
  publisher = {American Physical Society},
  doi = {10.1103/spng-p6c9},
  url = {https://link.aps.org/doi/10.1103/spng-p6c9}
}

@article{Bernadotte_Origin_2012,
    author = {Bernadotte, Stephan and Atkins, Andrew J. and Jacob, Christoph R.},
    title = {Origin-independent calculation of quadrupole intensities in X-ray spectroscopy},
    journal = {The Journal of Chemical Physics},
    volume = {137},
    number = {20},
    pages = {204106},
    year = {2012},
    month = {11},
    issn = {0021-9606},
    doi = {10.1063/1.4766359},
    url = {https://doi.org/10.1063/1.4766359},

}

@article{Hassing_The_2004,
author = {Hassing, S. and Svendsen, E. Nørby},
title = {The correct form of the Raman scattering tensor},
journal = {Journal of Raman Spectroscopy},
volume = {35},
year = {2004},
number = {1},
pages = {87-90},
doi = {https://doi.org/10.1002/jrs.1105},
url = {https://analyticalsciencejournals.onlinelibrary.wiley.com/doi/abs/10.1002/jrs.1105}
}

@article{Andersson_Second_1992,
    author = {Andersson, Kerstin and Malmqvist, Per‐Åke and Roos, Björn O.},
    title = {Second‐order perturbation theory with a complete active space self‐consistent field reference function},
    journal = {The Journal of Chemical Physics},
    volume = {96},
    number = {2},
    pages = {1218-1226},
    year = {1992},
    month = {01},
    issn = {0021-9606},
    doi = {10.1063/1.462209},
    url = {https://doi.org/10.1063/1.462209}
}
\end{document}